\RequirePackage{fix-cm}
\documentclass[smallextended]{svjour3}       % onecolumn (second format)
\smartqed  % flush right qed marks, e.g. at end of proof
\usepackage[english]{babel}
\usepackage[utf8x]{inputenc}
\usepackage[T1]{fontenc}
\usepackage{ulem,color}

\usepackage{amsmath}
\usepackage{amssymb}
\usepackage{amsfonts}
\usepackage{graphicx}
\usepackage[center]{caption}
\usepackage[colorlinks=true, allcolors=blue]{hyperref}

\begin{document}
	
	\title{Conformal invariance and vector operators in the $O(N)$ model}

	\author{Gonzalo De Polsi \and Matthieu Tissier \and Nicol\'as Wschebor
	}
	
	\institute{G. De Polsi \at
		Instituto de F\'isica, Facultad de Ciencias, Universidad de la
		Rep\'ublica, Igu\'a 4225, 11400, Montevideo, Uruguay \\
		\email{gdepolsi@fisica.edu.uy}           %  \\
		\and
		M. Tissier \at
		Sorbonne Universit\'e, CNRS, Laboratoire de Physique Th\'eorique de
		la Mati\`ere Condens\'ee, LPTMC, F-75005 Paris, France\\
		\email{tissier@lptmc.jussieu.fr}
		\and
		N. Wschebor \at
		Instituto de F\'isica, Facultad de Ingenier\'ia, Universidad de la
		Rep\'ublica, J.H.y Reissig 565, 11000 Montevideo, Uruguay\\
		\email{nicws@fing.edu.uy}
	}
	
	\date{Received: \today}
	
	\maketitle
	
	\begin{abstract}
		It is widely expected that, for a large class of models, scale
		invariance implies conformal invariance.  A sufficient condition for
		this to happen is that there exists no integrated vector operator,
		invariant under all internal symmetries of the model, with scaling
		dimension $-1$. In this article, we compute the  scaling dimensions of vector operators with  lowest dimensions in the $O(N)$ model. We use three different approximation schemes: $\epsilon$ expansion, large $N$ limit
		and third order of the Derivative Expansion of
		Non-Perturbative Renormalization Group equations. We find that the
		scaling dimensions of all considered integrated vector operators are
		always much larger than $-1$. This strongly supports the existence of conformal invariance in this model. For the Ising model, an argument based on  correlation functions inequalities was derived, which yields a lower bound for the scaling dimension of the vector perturbations. We generalize this proof to the case of the $O(N)$
		model with $N\in \left\lbrace 2,3,4 \right\rbrace$. 
	\end{abstract}
	
	\section{Introduction}
	
	Renormalization Group (RG) is a very efficient tool to study scale invariance and its consequences in critical phenomena: 
	Under mild assumptions, a fixed
	point of the RG transformation can be associated
	with a scale-invariant (critical) behaviour \cite{Wilson:1973jj}.  A modern version of
	Wilson's RG, often denoted as ``Non-perturbative Renormalization
	Group'' (NPRG), has been developed in the '90s
	\cite{Wetterich:1992yh,Ellwanger:1993kk,Morris:1993qb} (see \cite{Berges:2000ew} for a review; see \cite{Delamotte:2007pf} for
	a pedagogical introduction).  It allows to implement various
	kinds of approximations that, in some cases, can go far beyond
	perturbation theory. The typical situation encountered in such truncations (see for example,
	\cite{ZinnJustin:2002ru,Morris:1994ie,Morris:1994jc}) is that RG equations admit a discrete set of fixed points. By studying the characteristics of these fixed points as well as the RG flow around them, one can deduce the various universal quantities, such as critical exponents, scaling functions, etc.   In this context, it has been observed that the
	solution of the RG fixed point together with certain regularity
	properties of the generating functional of correlation functions completely
	characterizes all critical correlation functions at long distances
	(critical exponents, scaling functions, etc).  Unfortunately, solving the exact RG equations is beyond reach because these are 
	non-linear functional equations.  It is however reasonable to assume that the characteristics of the RG fixed points described above (existence of a discrete set of regular fixed points) are shared by the exact solutions.
	
	Since no exact solution is at reach so far and since we have to resort to approximations, any complementary
	information or insight that can be brought to ease the task is
	welcome.
	One clue in this direction comes from conformal invariance: there are
	strong indications that many systems at their critical point are not only scale invariant but
	show the full conformal group of symmetries (see \cite{Nakayama:2013is} for a review). This larger group gives
	strong constraints on the critical properties of many universality
	classes
	\cite{polyakov1974nonhamiltonian,ferrara1973tensor,rattazzi2008bounding,El-Showk:2014dwa,simmons2015semidefinite}
	without having to solve exactly Wilson's or NPRG equations. In fact,
	since the seminal paper of Belavin, Polyakov and Zamolodchikov
	\cite{Belavin:1984vu}, conformal symmetry has played a major role in
	the resolution of many bi-dimensional critical phenomena. More
	recently, a renewed interest in conformal symmetry in dimensions
	larger than two took place due to the successes of the
	``conformal bootstrap'' program.  The main idea is to implement an
	efficient algorithm which takes into account many constraints coming
	from conformal invariance, unitarity and crossing symmetry
	\cite{rattazzi2008bounding}. These constraints imply rigorous bounds on
	critical exponents that turn out to be impressively predictive in the
	$3d$ Ising universality class
	\cite{ElShowk:2012ht,El-Showk:2014dwa,Kos:2014bka}. The procedure has
	also been extended to other models as, for example, $O(N)$ invariant
	ones, but the constraints obtained in those cases are not, for the moment, as precise
	as in the Ising case \cite{Kos:2013tga,Kos:2015mba,Kos:2016ysd}.
	
	The success of the conformal bootstrap program re-opens the old
	question of determining whether a given model is conformal invariant
	or not in its critical regime. In fact, as early as the '70s it was postulated
	that many critical model could be conformal invariant
	\cite{Polyakov:1970xd,Migdal:1971xh}. However, at odds with scale
	invariance that was explained as a general property of RG equations,
	the origin and domain of validity of the invariance under conformal
	transformations is more complex and remained unclear until the proof
	by Zamolodchikov of its validity in the bi-dimensional case
	\cite{Zamolodchikov:1986gt}.\footnote{This proof has been criticized in the mathematical community (see, for example, \cite{Morinelli2019}). These observations apply
		also to the Polchinski's work \cite{polchinski1988scale} and to
		our previous \cite{delamotte2016scale} and present work. In the present paper, as in most theoretical physics studies, we
		do not pretend to achieve an analysis with the level of rigor of the mathematical community.} For larger dimensions, the issue was
	analyzed by Polchinski \cite{polchinski1988scale} in the '80s.
	More recently, important progress have been done in the analysis of the four dimensional case (see, for example, \cite{Luty:2012ww,Dymarsky:2013pqa,Dymarsky:2014zja,Dymarsky:2015jia}). In general dimensions, including the three-dimensional case, Polchinski
	showed that under a certain sufficient condition, scale invariance
	implies conformal invariance. More recently, a similar sufficient
	condition was derived in the framework of NPRG equations
	\cite{delamotte2016scale}.  These two sufficient conditions both assume that, in the critical 
	regime, the model is invariant under translations and rotations. On top
	of this assumption the sufficient condition in
	\cite{polchinski1988scale} assumes that interactions are sufficiently
	short-ranged in order to ensure the existence of a local
	energy-momentum tensor with standard properties. In the case of
	\cite{delamotte2016scale}, it is required that the RG flow around the fixed point is sufficiently regular. Finally, both conditions require some information on the scaling
	dimensions of  operators which
	transform as vectors under space translations and rotations and which
	are invariant under all internal
	symmetries of the universality class under consideration. We shall call such operators "vector operators" from now on.
	The sufficient condition of \cite{polchinski1988scale} requires that there exist no local vector operator with scaling dimension $d-1$ (apart from possible total derivatives). The sufficient condition of   \cite{delamotte2016scale} instead focus on {\it integrated} operators and requires that there exists no such operators of dimension $-1$. 
	
	A natural path to prove that conformal invariance is indeed realized
	in a given model is therefore to compute the lowest scaling dimension
	of vector operators or find a lower bound for that quantity. In~\cite{delamotte2016scale} it was proven under some assumptions,
	and using inequalities on correlation functions
	\cite{griffiths1967correlationsI,kelly1968general,lebowitz1974ghs},
	that all local vector operators have scaling dimensions strictly
	larger than $d-1$ for the Ising universality class.
	Accordingly, in this important universality class, scale invariance
	implies conformal invariance in all dimensions. This proof has been
	criticized in \cite{Meneses:2018xpu} where it was
	argued that the assumptions made may not be fulfilled.  Some elements of reply have already
	been presented in \cite{Delamotte:2018fnz} but we discuss below in
	detail the issues raised in~\cite{Meneses:2018xpu}. In~\cite{delamotte2016scale} the scaling
	dimension of the most relevant vector operator was also calculated in
	$d=4-\epsilon$ in the Ising universality class obtaining
	$D_V=3+\mathcal{O}(\epsilon^2)$.  The most relevant integrated
	operator near $d=4$ has the form
	$\int d^dx\,\phi^3\partial_\mu\partial^2\phi$.  In~\cite{Meneses:2018xpu} the scaling dimension of the same operator
	(modulo a total derivative) was analyzed in a 3D Monte-Carlo simulation
	obtaining $D_V=3\pm 1$. 
	
	In the present article, we extend the analysis of the scaling
	dimensions of vector operators to $O(N)$ models. We compute the  scaling dimensions by three different approximation
	schemes. First, we review in detail the $d=4-\epsilon$ calculation
	for the scaling dimensions of leading vector operators (already given in \cite{delamotte2016scale}). Second,
	we perform the same calculation in the large $N$ limit for any
	dimension $d$.  Third, we calculate the scaling dimension of various
	vector operators (including leading ones) by using the Derivative
	Expansion (DE) of the NPRG. The
	overall picture is that the estimates coming from the three
	approximation schemes coincide with a high level of precision. This
	allows us to conclude that the most relevant integrated vector
	operator has scaling dimension larger than 2 for any $d\ge 3$ and any $N\ge 1$, well above the bound $-1$ appearing in the sufficient condition.
	
	Among the different eigenoperators which transform as vectors that are found in the three methods described above, some of them happen to be proportional to the equations of motion, see below for details. Such operators are called redundant and are often considered as non-physical because they can be reabsorbed through a change of variables. In the specific context we are interested in, these redundant operators are indeed not physically relevant in the precise sense that if there would exist an integrated redundant vector operator of dimension -1, it would not be a possible source of breaking of conformal invariance, as we shall show below. However, in the NPRG framework, reabsorbing redundant operators is often not particularly convenient, see however \cite{Gies:2001nw}. We therefore consider {\it all} vector eigenoperators, may they be redundant or not. By constraining the scaling dimensions of all vector operators, we constraint in particular the non-redundant ones, which are the only ones which may lead to a breaking of conformal invariance. Redundant operators are also particular because it is sometimes possible to compute their scaling dimension easily. We show that some of these operators have indeed simple scaling solutions. Knowing a priori the scaling dimension of some operators is valuable when analysing the scaling dimensions obtained in the three different analytical approaches considered in this article.

	On top of computing the scaling dimensions of vector operators in different approximate schemes, we give a lower bound for
	the scaling dimensions $D_V$. We follow the strategy used in
	\cite{delamotte2016scale}, and extend it to the $O(N)$ model.
	This work is based on known generalizations of Griffiths and
	Lebowitz inequalities that are valid for the $O(N)$ model (for
	$N=2, 3,$ and $4$)
	\cite{bricmont1977gaussian,dunlop1979zeros,dunlop1976correlation,monroe1979correlation,kunz1975correlation,sokal1982mean}.

	The article is organized as follows. In Sect.~\ref{secNPRG}, we
	review shortly the $O(N)$ models and NPRG equations. We then recall in  Sect.~\ref{Sec.II}  the sufficient condition under which scale invariance
	implies conformal invariance. In Sect.~\ref{Sec.V}, we compute the
	scaling dimension of the most relevant vector operators in the $\epsilon$ expansion, the large
	$N$ expansion and the
	$\mathcal{O}(\partial^{3})$ approximation of the DE in the $O(N)$ case
	for different values of $N$. The proof for the lower bound of $D_V$ in
	the Ising case is reviewed in Sect.~\ref{Sec.III}. We give some
	details that were omitted in the previous paper
	\cite{delamotte2016scale}. Sect.~\ref{Sec.IV} deals with the
	extension of the proof from the Ising model to the $O(N)$ for
	$N\in \left\lbrace 2,3,4 \right\rbrace$. We give our
	conclusions in Sect.~\ref{Sec.VI}. We discus  in the Appendix~\ref{Ap:redundant} the issues related to
	redundant operators, which are not central for our main line of arguments, and give some detail of the calculations in appendices \ref{Ap:EpsExp}, \ref{Ap:LargeN} and \ref{Ap:NumParam}. 
	
	\section{NPRG and scale invariance}\label{secNPRG}

	\subsection{O(N) models}
	We consider a model
	with $N$ scalar fields $\varphi_i$ with $i=1,\dots N$ in a $d-$dimensional euclidean space. We choose as a Hamiltonian (or Euclidean action) the standard Ginzburg-Landau $\varphi^4$ model:
	\begin{equation}
	\label{action}
	S[\varphi]=\int_x \Big\{\frac 1 2 \partial_\mu\varphi_i\partial_\mu\varphi_i+\frac r 2 \varphi_i\varphi_i +\frac{u}{4!}(\varphi_i\varphi_i )^2\Big\}
	\end{equation}
	where $\int_x=\int d^dx$. Here and below, Einstein convention is
	employed both for internal indices $i$ and for space indices $\mu$,
	unless otherwise stated.\footnote{For $N=1$, the theory describes a single
		scalar with $\mathbb{Z}_2$ symmetry.} We will also consider the
	analytic extension to values of $N$ that are not positive integers. In
	particular, we will study the $N=0$ case which is relevant for the
	problem of self-avoiding polymer chains \cite{degennes72}. 
	
	We note that in several physical systems belonging to the $O(N)$ universality class, the microscopic action
	is {\it not} $O(N)$ invariant: the $O(N)$ symmetry is an
	emergent phenomenon near the critical point. For
	simplicity, we do not consider this possibility in this article and assume that
	the microscopic action is invariant under the $O(N)$ symmetry.
	
	\subsection{NPRG equations}
	
	The NPRG is based on Wilson's ideas of integrating first the highly
	oscillating modes (i.e. those with a wavevector larger than some
	scale $k$) while keeping untouched the long-distance modes.
	A convenient implementation consists in adding to the action a
	regulating term  quadratic in the fields \cite{Polchinski:1983gv},
	$S[\varphi]\to S[\varphi]+\Delta S_k[\varphi]$ with:
	\begin{equation}
	\label{deltaS}
	\Delta S_k[\varphi]=\frac 1 2 \int_{x,y}\varphi_i(x) R_k(x,y) \varphi_i(y).
	\end{equation}
	The regulating function $R_k$ is chosen to be invariant under
	rotations and translations and therefore depends only on
	$|x-y|$. Moreover, its Fourier transform $R_k(q)$ should
	\begin{itemize}
		\item be a smooth function of the modulus of the momentum $q$;
		\item behave as a ``mass'' of order $k$ for long-distance modes:
		$R_k(q)\sim Z_k k^2$ for $q\ll k$, where $Z_k$ is a field renormalization factor to be specified below;
		\item go to zero rapidly when $q\gg k$ (typically faster than any power law).
	\end{itemize}
	With these properties the term (\ref{deltaS}) regularizes the theory
	in the infrared without modifying the ultraviolet regime. One can then
	define a scale-dependent generating functional of connected
	correlation functions \cite{Wetterich:1992yh,Ellwanger:1993kk,Morris:1993qb}:
	\begin{equation}
	\label{regulatedgeneratingfunc}
	e^{W_k[J]}=\int \mathcal{D}\varphi\  e^{-S[\varphi]-\Delta S_k[\varphi]+\int_x J_i(x) \varphi_i(x)},
	\end{equation}
	and a scale-dependent effective action defined as the modified Legendre transform:
	\begin{equation}
	\Gamma_k[\phi]=\int_x \phi_i(x) J_i(x) -W_k[J]-\Delta S_k[\phi].
	\end{equation}
	In the previous equation, $J$ is an implicit function of $\phi$,
	obtained by inverting
	\begin{equation}
	\phi_i(x)=\frac{\delta W_k}{\delta J_i(x)}.
	\end{equation}
	The running of $\Gamma_k[\phi]$ with the RG time $t=\log(k/\Lambda)$  \cite{Wetterich:1992yh,Ellwanger:1993kk,Morris:1993qb} can be easily obtained:
	\begin{equation}
	\partial_{t}\Gamma_{k}[\phi]=\frac{1}{2}\int_{x,y}\partial_{t}R_{k}(x-y)G_k(x,i;y,i)
	\label{wettericheq}
	\end{equation}
	Here $G_k(x,i;y,j)$ is the propagator in an external field, which has a matrix structure because of the vector indices:
	\begin{equation}
	\int_{y} G_k(x,i;y,n)\left[\frac{\delta^2\Gamma_k}{\delta \phi_n(y)\delta\phi_j(z)}+ R_k(y-z)\delta_{nj}\right]=\delta(x-z)\delta_{ij}
	\end{equation}
	
	The main advantage of this version of the RG with respect to the more standard Wilson \cite{Wilson:1973jj} or Polchinski \cite{Polchinski:1983gv} RG equation is that the right hand side. only includes
	1PI dressed diagrams (to be compared to Polchinski equation where 1PR connected diagrams contribute also). This 1PI property makes the equations
	much better suited for the formulation of approximations that go beyond perturbation theory (see, for example \cite{Berges:2000ew}). 
	
	\subsection{Scale invariance in the NPRG context}

	In this section, we recall how scale invariance is treated in Wilson's RG. The more involved case of special conformal transformations is discussed below. 
	
	Scale invariance emerges in the NPRG (as in Wilson's RG) due to the
	existence of a fixed point for renormalized, dimensionless
	quantities. These are defined by introducing:
	\begin{align}
	\tilde x&=k x\\
	\tilde q&=k^{-1} q\\
	\tilde \phi_i(\tilde x)  &=k^{-(d-2)/2}Z_k^{1/2}\phi_i(x).
	\end{align}
	where $Z_k$ is the field-renormalization factor. In terms of
	dimensionless and renormalized variables, the NPRG flow equation (\ref{wettericheq}) becomes:
	\begin{equation}
	\partial_t \Gamma_k[\tilde \phi]=\int_{\tilde x} \frac{\delta \Gamma_k}{\delta \tilde \phi_i(\tilde x)}\left(D^ {\tilde x}
	+D_k^{\phi}\right)\tilde \phi_i(\tilde x)+\frac 12\int_{\tilde x\tilde y}\partial_t\tilde R(\tilde x-\tilde y) \tilde G_{k}(\tilde x,i;\tilde y,i)
	\label{eq_flow}
	\end{equation}
	where $R_k( x)=Z_k k^{d+2}\tilde R(k x)$, $D_k^{\phi}=(d-2+\eta_k)/2$, $D^x=x_\mu\partial_{x^\mu}$ and $\eta_k=-\partial_t \log(Z_k)$.\footnote{At a fixed point, the running anomalous dimension $\eta_k$ identifies with the anomalous dimension $\eta$ which governs the decay of correlation functions at long distances.} The field renormalization factor $Z_k$
	can be fixed in many ways. A convenient prescription consists in choosing
	$Z_k$ by imposing a normalization condition of the field compatible
	with the tree-level action, for instance:
	\begin{equation}
	\left. \frac{\partial}{\partial \tilde p^2}\Bigg(\int_x e^{i \tilde p\cdot (\tilde x-\tilde y)}
	\frac{\delta^2\Gamma_k}{\delta \tilde \phi_i(\tilde x)\delta\tilde \phi_j(\tilde y)}\Bigg)\right|_{\tilde p=0}
	=\delta_{ij}
	\end{equation}
	Note that the theory is IR-regularized thanks to the addition of
	$\Delta S_k$, see Eq.~(\ref{deltaS}). The previous derivative is
	therefore well-defined.  For $O(N)$ models it is well established that
	the dimensionless and renormalized flow equation has a {\it fixed
		point solution} (at least for dimensions
	close enough to $d=4$):
	\begin{equation}
	\partial_t \Gamma_*[\tilde \phi]=0
	\label{eq_fixedpoint}
	\end{equation}
	This fixed point condition can be re-expressed in terms of the
	dimensionfull field and it reads:
	\begin{equation}
	\int_{x} \frac{\delta \Gamma_*}{\delta \phi_i(x)}\left(x^\rho\partial_{x^\rho}
	+D_*^{\phi}\right)\phi_i(x)+\frac 12\int_{x y}\partial_t R_k(x-y) G_{*}(x,i;y,i)=0
	\label{fixedpointcondition}
	\end{equation}
	where $D_*^{\phi}$ is the fixed point value of the scaling dimension of the field and $G_{*}(x,i;y,j)$ is the dimensionfull exact propagator in presence of an external field.

	The condition (\ref{fixedpointcondition}) has a simple interpretation
	as a Ward identity for scale invariance of the fixed point
	solution. Indeed, it can be shown that the second term of
	Eq.~(\ref{fixedpointcondition}) vanishes for the modes with wave numbers
	much larger than $k$.\footnote{This property, called ``decoupling''
		in~\cite{Canet:2014dta}, relies on the fact that correlation
		functions in non-exceptional configurations of momenta (or
		distances) have a finite limit when $k\to 0$.  In the case of
		turbulence, this property of decoupling is not fulfilled. This
		results in the fact that correlation functions are power-laws as in scale invariant theories,
		but the critical exponents for n-point correlation functions are not
		simply related one with another. This property is usually called
		multifractality (see, for example, \cite{Frisch1995}).} Suppose for a moment that the microscopic action is chosen
	such that Eq.~(\ref{fixedpointcondition}) is fulfilled for all
	$k$, then Eq.~(\ref{fixedpointcondition})
	states that the effective action, and therefore all correlation
	functions, are invariant under dilatations (modulo terms that regularize the Ward identity in the infrared). Of course, since we can
	make $k$ as small as wanted, this implies that scale invariance is
	valid for all modes. Note that, the present analysis only works if we assign the right scaling to the regulator, $R_k(x-y)=Z_k k^{d+2}f(k(x-y))$ with $Z_k\propto k^{-\eta}$ near the fixed point. 
	
	As noted in \cite{delamotte2016scale}, the Eq.~(\ref{fixedpointcondition}) can be interpreted in another way.
	Instead of varying the fields $\phi_i$ at fixed regulator $R_k$, one can consider the regulator $R_k(x-y)$ as a bi-local external source that can be varied covariantly with respect to
	scale transformations. Scale transformation then consists in the simultaneous variations:
	\begin{align}
	\label{variationsscale}
	\delta\phi_i(x)&=(D^x+D_*^\phi)\phi_i(x)\nonumber\\
	\delta R_k(x-y)&=(D^x+ D^y +D_*^R)R_k(x,y)
	\end{align}
	where  $D_*^R=2d-2D_*^\phi$ is the scaling dimension of $R_k$.
	With this in mind one can interpret (\ref{fixedpointcondition}) as a Ward-Identity for scale-invariance of $\Gamma_*$ where the field $\phi_i$ and $R_k(x-y)$ are transformed simultaneously:
	\begin{equation}
	\label{idWardmoddilat-R}
	\int_{xy} (D^x+ D^y +D_*^R)R_k(x,y)\frac{\delta \Gamma_*}{\delta R_k(x,y)}
	+\int_{x} (D^x +D_*^\phi)\phi(x)\frac{\delta \Gamma_*}{\delta \phi(x)}=0
	\end{equation}
	
	In practice, though, invariance under dilatations is not valid for
	distances comparable with the microscopic scale. There are two reasons for that.
	First, typically the microscopic theory has an underlying scale and the microscopic
	action is usually {\it not} scale invariant. Second, even when the microscopic action
	is scale invariant, this classical symmetry can be broken by anomalies (see, \cite{Morris:2018zgy} for an analysis of the trace anomaly in the NPRG context).
	Instead of being present at the microscopic level, scale
	invariance usually appears as an emergent property. This is easily understood
	as follows. By fine tuning the initial condition of the flow, one
	obtains a RG trajectory which asymptotically approaches the fixed
	point in the infrared. If we denote $t_G=-|t_G|$ the typical RG ``time''
	necessary to reach the vicinity of the fixed point, we conclude that
	scale invariance occurs for length scales larger than the so-called
	Ginzburg length $l_G=\Lambda^{-1} e^{-t_G}$. For larger RG times (in absolute
	value), the running effective action is close to the fixed point and
	the condition Eq.~(\ref{eq_fixedpoint}) is approximately fulfilled. Of
	course, we can choose $k$ as small  as we like (or equivalently $|t|$ as large as we like) so that the fixed point
	condition (and consequently, the Ward identity for dilatations) is
	fulfilled with arbitrary precision in the long-distance regime.
	
	A similar discussion can be done if the system is not exactly critical
	({\it i.e.} if the initial condition is not exactly fine tuned). The
	RG flow is then divided in 3 regimes. For $|t|<|t_G|$, the flow drives
	the system close to the fixed point. For larger RG times, the flow is
	then very slow and the fixed-point condition Eq.~(\ref{eq_fixedpoint})
	is again approximately fulfilled. At a RG scale $k\sim \xi^{-1}$, where
	$\xi$ denotes the correlation length, the system starts to depart
	exponentially fast in the RG time from the fixed point. In this situation, there is a
	regime of momenta $\xi^{-1}\ll p\ll l_G^{-1}$ for which the theory is
	approximately scale invariant.
	
	Apart from the fixed point effective
	action $\Gamma_\star[\tilde \phi]$, from which we can deduce the
	anomalous dimension $\eta$, there are other quantities of
	interest, which are related with universal observables. These
	are obtained by considering a small perturbation around the fixed
	point:
	$\tilde{\Gamma}[\tilde{\phi}]=\tilde{\Gamma}^{*}[\tilde{\phi}]+\varepsilon
	e^{\lambda t} \tilde{\gamma}[\tilde{\phi}]$. Expanding at linear order in
	$\varepsilon$, we obtain the eigenvalue equation:
	\begin{align}
	&\lambda \tilde{\gamma}[\tilde{\phi}]=\int_{\tilde{x}}(D^{\tilde{x}}+D_{\phi})\tilde{\phi_i}(\tilde{x})\frac{\delta\tilde{\gamma}}{\delta\tilde{\phi_i}(\tilde{x})}\nonumber\\
	&-\frac{1}{2}\int_{\tilde{x_i}}[(D^{\tilde{x}}+D_R)\tilde{R}(|\tilde{x}-\tilde{y}|)]\tilde{G}^{*}(\tilde{x},i;\tilde{z},j)\tilde{\gamma}^{(2)}(\tilde{z},j;\tilde{w},k)\tilde{G}^{*}(\tilde{w},k;\tilde{y},i)
	\label{eq.perturb}
	\end{align}
	\noindent where
	$\tilde{x}_i=\lbrace\tilde{x},\tilde{y},\tilde{z},\tilde{w}\rbrace$,
	and
	$\tilde{G}^{*}=\left({\tilde{\Gamma}}^{*(2)}+\tilde{R}\right)^{-1}$.  The
	spectrum of eigenvalues $\lambda$ is expected to be discrete,\footnote{The quantization of the eigenvalue spectrum is associated with the fact that we must only consider perturbations $\tilde{\gamma}[\tilde{\phi}]$ which lead to correlation functions which are regular at long distances (small momenta) in the presence of the infrared regulator. Indeed, the initial condition of the flow involves correlation functions which can be Fourier transformed and are infinitely differentiable with respect to fields and wavevectors and this property is preserved by the flow.} although
	no proof of this is known at the level of the exact equation. This has been very thoroughly studied in perturbation
	theory (see, for example, \cite{ZinnJustin:2002ru}) and also within the Derivative Expansion of NPRG equations (see \cite{Morris:1994ie,Morris:1994jc}, for example). At the Wilson-Fisher fixed point, there
	exists one negative eigenvalue associated with an $O(N)-$invariant eigenvector,
	which corresponds to the fact that we need to
	fine tune only one parameter (say the temperature) to reach the 
	criticality at zero external magnetic field. This negative eigenvalue is directly related to the
	critical exponent $\nu$ which governs the divergence of the
	correlation length at criticality. The positive eigenvalues encode the
	correction to scaling exponents.
	
	Equation (\ref{eq.perturb}) can also be rewritten by considering a simultaneous variation of $\phi_i$ and $R_k(x-y)$ under scale transformations (see Eq.~(\ref{variationsscale})):
	\begin{align}
	\mathcal{D}\gamma[\phi]&\equiv \int_{xy} (D^x+ D^y +D_*^R)R_k(x,y)\frac{\delta \gamma}{\delta R_k(x,y)}
	+\int_{x} (D^x +D_*^\phi)\phi(x)\frac{\delta \gamma}{\delta \phi(x)}\nonumber\\
	&=\lambda \gamma[\phi]
	\end{align}
	This equation can be interpreted as an eigenproblem of the dilatation operator where, as in Eq.~(\ref{fixedpointcondition}), the operator acts on the field $\phi$ and the regulator $R$.
	
	\section{A sufficient condition for conformal invariance}\label{Sec.II}
	
	Let us now discuss conformal invariance following the same
	line of arguments as for scale invariance. Assuming for a moment that the microscopic action and the path
	integral measure are conformal invariant, the Ward identity for
	conformal invariance can be expressed as \cite{delamotte2016scale} (for a similar analysis of conformal invariance in the NPRG context, see
	\cite{Rosten:2014oja,Rosten:2016zap}):
	\begin{equation}
	\label{eq_cond_conforme}
	\Sigma_{k}^{\mu}[\phi]=0,
	\end{equation}
	where $\Sigma_{k}^{\mu}[\phi]$ is defined as:
	\begin{equation}
	\label{def_sigma}
	\Sigma_{k}^{\mu}[\phi]\equiv  \int_{x} (K_{\mu}^{x}-2D_\star^{\phi}x_{\mu})\phi_i(x)\frac{\delta\Gamma_{k}}{\delta\phi_i(x)}-\frac{1}{2}\int_{x,y}\partial_{t}R_{k}(|x-y|)(x_{\mu}+y_{\mu})G_{k}(x,i;y,i)
	\end{equation}
	and
	$K_{\mu}^{x}=x^2\partial_{\mu}^{x}-2x_{\mu}x_{\nu}\partial_{\nu}^{x}$. From its definition, we easily find that $\Sigma_{k}^{\mu}$ is a scalar under O($N$) transformations (more generally, under internal transformations) and that the associated dimensionless quantity should be defined as 
	$\tilde{\Sigma}_{k}^{\mu}[\tilde{\phi}]=k\Sigma_{k}^{\mu}[\phi]$, the flow of which can be obtained by using the exact flow equation (\ref{wettericheq}).
	A straightforward but lengthy calculation leads to \cite{delamotte2016scale}:
	\begin{align}
	\partial_{t}\tilde{\Sigma}_{k}^{\mu}[\tilde{\phi}]&-\tilde{\Sigma}_{k}^{\mu}[\tilde{\phi}]=
	+\int_{\tilde{x}}(D^{\tilde{x}}+D_\star^{\phi})\tilde{\phi_i}(\tilde{x})\frac{\tilde{\delta\Sigma_{k}^{\mu}}}{\delta\tilde{\phi_i}(\tilde{x})} \nonumber\\
	&-\frac{1}{2}\int_{\tilde{x}_i}[(D^{\tilde{x}}+D_\star^R)r(\tilde{x}-\tilde{y})]\tilde{G}_{k}(\tilde{x},i;\tilde{z},j)\tilde{\Sigma}_{k}^{\mu(2)}(\tilde{z},j,\tilde{w},k)\tilde{G}_{k}(\tilde{w},k,\tilde{y},i)
	.\label{eq.Conform}
	\end{align}
	At the fixed point, Eq.~\eqref{eq.Conform} coincides with the
	eigenvalue problem Eq.~\eqref{eq.perturb} (with eigenvalue
	$\lambda=-1$) if we replace $\tilde{\gamma}[\tilde{\phi}]$ by
	$\tilde{\Sigma}_{\star}^{\mu}[\tilde{\phi}]$. This means that, if it exists, $\tilde{\Sigma}_{k}^{\mu}$ is a vector eigenoperator with dimension -1.

	As for scale invariance, the Ward identity for conformal invariance  can be reinterpreted in terms of simultaneous variations
	of fields $\phi_i(x)$ and the regulator $R_k(x-y)$ under
	special conformal transformations:
	\begin{align}
	\label{variationsconf}
	\delta\phi_i(x)&=(K_{\mu}^x-2 D_\star^\phi x_\mu)\phi_i(x)\nonumber\\
	\delta R_k(x-y)&=(K_{\mu}^x-\, D_\star^R\, x_\mu+ K_{\mu}^y-
	D_\star^R\, y_\mu)R_k(x,y)
	\end{align}
	Indeed, a simple calculation leads to the equality:
	\begin{equation}
	\begin{split}
	\Sigma_k^\mu[\phi]=\mathcal K_\mu \Gamma_k\equiv&\int_x (K_{\mu}^x-2 D_\star^\phi x_\mu)\phi(x)\frac{\delta \Gamma_k}{\delta \phi(x)}\\
	+\int_{xy} (&K_{\mu}^x-\, D_\star^R\, x_\mu+ K_{\mu}^y-
	D_\star^R\, y_\mu)R_k(x,y)\frac{\delta \Gamma_k}{\delta R_k(x,y)},
	\end{split}
	\end{equation}
	In fact, similar expressions
	can be obtained for the generators of translations $\mathcal P_\mu$
	and rotations $\cal J_{\mu\nu}$:\footnote{The regulator has been chosen to depend only on $|x-y|$ and is therefore invariant under translations and rotations. These transformations therefore involve no variation of the regulator.}
	\begin{equation}
	\mathcal P_\mu \Gamma_k=\int_x\, \partial_\mu\phi(x)\frac{\delta \Gamma_k}{\delta \phi(x)}
	\end{equation}
	\begin{equation}
	\mathcal J_{\mu\nu} \Gamma_k=\int_x\, (x_\mu \partial_\nu\phi(x)\
	-x_\nu \partial_\mu\phi(x))
	\frac{\delta \Gamma_k}{\delta \phi(x)}
	\end{equation}
	It can easily be checked that the generators satisfy the algebra of the conformal group. In particular, applying $[\mathcal P_\mu,\mathcal K_\nu]=2\delta_{\mu\nu}\mathcal D-2\mathcal J_{\mu\nu}$ to a translation, rotation and
	dilatation invariant $\Gamma_*$ yields 
	\begin{equation}
	\label{eq_inv_trans_sigma}
	\mathcal P_\mu \Sigma_*^\nu=0.  
	\end{equation}
	Thus, $\Sigma_*^{\mu }$ is the integral of a local vector functional (in the sense that it is a function of the field and its derivatives at a given point, with no explicit dependence on the position) \cite{delamotte2016scale}. Similarly, by applying the commutator $[\mathcal J_{\mu\nu},\mathcal K_\rho]$ to the fixed point $\Gamma_\star$, we readily find that, if it exists, $\Sigma_\star^{\mu}$ transforms as a vector under rotations.
	
	We can also use the conformal algebra to re-derive the fact that $\Sigma_*^{\mu}[\phi]$, if it exists, must be an eigenfunction of the dilation operator with eigenvalue $-1$. Indeed, applying the commutation relation
	$[\mathcal{D},\mathcal{K}_\mu]=-\mathcal{K}_\mu$ to the scale-invariant $\Gamma_*$ (which satisfies $\mathcal{D}\Gamma_*=0)$ leads to:
	\begin{equation}
	\mathcal{D}\Sigma_*^\mu=-\Sigma_*^\mu.
	\end{equation}
	This, again, implies that, if it exists, $\Sigma_*^\mu$ is an eigenvector of $\mathcal{D}$ with scaling dimension $-1$.
	
	To summarize, we found that, if it exists, $\Sigma_*^\mu$  is an eigenvector of the dilatation operation with eigenvalue $-1$, is invariant under translations and internal symmetries [O($N$) here] and transforms like a vector under rotations. The sufficient condition is easily derived now. Indeed, suppose that in a given model there
	exists no eigenvector with the aforementioned properties. Under these circumstances, the only
	solution to Eq.~(\ref{eq.Conform}) is $\Sigma_*^\mu=0$, which
	implies that conformal invariance is fulfilled at the fixed point, see
	Eqs.~(\ref{eq_cond_conforme},\ref{def_sigma}).
	
	It is important to relate the present analysis to the previous analysis of Polchinski \cite{polchinski1988scale}. In fact, for models where interactions are sufficiently short-range, instead of analyzing Ward identities, one can consider the associated Noether theorem with the corresponding currents. In that case, all the present analysis can be cast in terms of densities of the various quantities and, in particular, the density of $\Sigma_*^\mu$ is called the virial current (that, by construction, is defined modulo a total derivative) and instead of requiring that there is no integrated vector operator of dimension $-1$ one must require that there is no local vector operator with dimension $d-1$. In that aspect both analysis are equivalent. Let us note, however, that the present analysis, formulated in terms of integrated operators is more general and includes situations where the interactions are long-range, such as the Ising model with an exchange term decreasing as $1/r^{d+\sigma}$ with $\sigma>0$ \cite{PhysRevLett.29.917}.
	
	%@article{PhysRevLett.29.917,
	%  title = {Critical Exponents for %Long-Range Interactions},
	%  author = {Fisher, Michael E. and %Ma, Shang-keng and Nickel, B. G.},
	%  journal = {Phys. Rev. Lett.},
	%  volume = {29},
	%  issue = {14},
	%  pages = {917--920},
	%  year = {1972},
	%}
	
	\section{Vector operators in the \textit{O(N)} model} \label{Sec.V}
	
	The sufficient condition described above gives a natural way to prove
	that the O($N$) model is conformal at criticality. It consists in
	computing the scaling dimensions of vector operators and check whether
	or not they are equal to $-1$. Unfortunately, computing critical
	exponents exactly in $d=3$ is notoriously difficult. We will thus use
	various approximation schemes. We first review the $4-\epsilon$
	calculation. We then consider the large-$N$ limit, and finally, the DE of the NPRG equations at order
	$\mathcal{O}(\partial^3)$.
	
	In all of these three approaches, the calculation follows the same
	steps. We add a small perturbation to the Wilson-Fisher fixed point with a set of vector
	operators $V_\mu^i$. This can be done by adding to the action terms
	of the form $a_\mu^iV_\mu^i$. Since the Wilson-Fisher fixed
	point corresponds to $a_\mu^i=0$,\footnote{Indeed, the Wilson-Fisher
		fixed point is rotational invariant but a non-zero value of
		$a^i_{\mu}$ would break isotropy.} we only need
	to retain the terms linear in $a_\mu^i$ in the beta functions
	$\beta_{a_\mu^i}$:
	$\beta_{a_\mu^i}=M_{ij}a_\mu^j+\mathcal O(a_\mu^i a_\nu^j)$. $M$ is
	the stability matrix in the vector sector. The scaling dimensions of the
	vector operators is then obtained by diagonalizing the matrix $M$.\footnote{Here and below we assume, as usual, that the matrix
		$M$ is diagonalizable.} We
	stress that, in this whole section, we consider integrated operators
	(as opposed to local ones). We can therefore use different
	parametrizations of a vector operator which differ by integration by
	parts without altering the final result. For instance,
	$\int _x \phi^3 \partial_\mu (\partial^2)\phi$ and
	$\int _x 3\phi \partial_\mu\phi (\partial\phi)^2$ are completely
	equivalent writings of the same quantity.

	Before going into the details, let us give the overall result:
	$4-\epsilon$ and DE at $\mathcal{O}(\partial^3)$ give similar
	results for any $d\ge 3$. This is an indication that for scaling dimensions of operators with three derivatives, the
	one-loop approximation is a good estimate in $d=3$. Even considering very
	pessimistic error bars (see below) the value $-1$ for an $O(N)$
	invariant integrated vector operator is unambiguously excluded. For dimensions lower than
	$\sim 2.5$, we are unable to control error bars, and the Wilson-Fisher fixed point is not under control
	at order $\mathcal{O}(\partial^3)$ (except for a small window in $d=2+\epsilon$). Many sources of instability appear: in some cases the $\mathcal{O}(\partial^0)$ fixed point becomes unstable and
	in other cases some critical exponents become complex. Therefore, we only present the results for $d\ge 2.5$ and only present
	the critical exponents when they are real. In any case, it is clear that the vector operators with complex scaling dimension are not candidates
	for potential sources of breaking of conformal symmetry but they are probably the indication of an uncontrolled behaviour of our approximations in low dimensions. One observes that the error bars can become large below $d=3$ but in the
	physically interesting case of $d=3$ they are always small and unambiguously exclude the $-1$ value.
	Given the previous results proving
	that for $O(N)$ models scale invariance implies conformal invariance in $d=2$ \cite{Zamolodchikov:1986gt,polchinski1988scale},
	one concludes that there is a very strong indication of the presence of conformal invariance in the
	critical regime of $O(N)$ models for any $N$ and any $d$. 
	
	\subsection{The $\epsilon$ expansion of vector scaling dimensions}
	\label{sec_eps}
	In $d=4$, there are two independent $O(N)-$invariant vector operators
	which have the lowest scaling dimensions. We therefore introduce the perturbation
	\begin{align}
	\label{eq_deltas}
	\delta S= \int d^{d}x \,\left[\frac{a_{\mu}}{4}\phi_i \partial^{\mu}\phi_i \partial_\nu\phi_j\partial_\nu\phi_j+\frac{b_{\mu}}{2}  \phi_i\partial_{\nu}\phi_i \partial_{\nu} \phi_j  \partial^{\mu}  \phi_j \right]
	\end{align}
	to the microscopic action.
	
	In order to compute the scaling dimension at leading order in
	$\epsilon=4-d$ we need to compute only the one-loop diagrams. As
	usual, we introduce dimensionless variables (that we denote below with
	the same symbols, for notation simplicity). The flow
	equations read:
	\begin{equation}
	\begin{split}
	\partial_{t}u&=-\epsilon u +\frac{\left(N+8\right)u^{2}J}{3} +\mathcal O(u^3,a^2,b^2,ab)\\
	\partial_{t}a_{\mu}&=\left(3-\epsilon\right)a_{\mu}+\frac{uJ}{3}\left[
	\left(N+4\right)a_{\mu}+4b_{\mu} \right]+\mathcal
	O(u^2a,u^2b,a^2,b^2,ab)\\
	\partial_{t}b_{\mu}&=\left(3-\epsilon\right)b_{\mu}+\frac{uJ}{3}\left[ 2a_{\mu}+\left(N+6\right)b_{\mu} \right]+\mathcal
	O(u^2a,u^2b,a^2,b^2,ab)    
	\end{split}
	\label{eq:feEps}
	\end{equation}
	\noindent where $J$ is the dimensionless version of
	$\int_{q}\dot{R_{k}}\left(q\right)G_{k}^{3}\left(q\right)$ with
	$G_k\left(q\right)$ the regularized propagator.\footnote{As is well
		known \cite{Berges:2000ew}, the integral $J=1/(16\pi^2)+\mathcal{O}(\epsilon)$, independently of the particular
		choice of
		the regulator $R_k$. This ensures the universality of the $\beta$
		functions given here.}  The (Wilson-Fisher)
	fixed point solution reads:
	\begin{equation}
	u^*=3\frac{\epsilon}{(N+8)J},\hspace{1cm} a^{*}_{\mu}=0, \hspace{1cm}b^{*}_{\mu}=0.
	\end{equation}
	Substituting the fixed point solution in Eq.~(\ref{eq:feEps}) we
	obtain the linearized flow for the couplings $a_{\mu}$ and $b_{\mu}$:
	\begin{equation}
	\begin{split}
	\partial_{t}a_{\mu}&=\left(3-\epsilon\right)a_{\mu}+\frac{\epsilon}{N+8}\left[ \left(N+4\right)a_{\mu}+4b_{\mu} \right]\\
	\partial_{t}b_{\mu}&=\left(3-\epsilon\right)b_{\mu}+\frac{\epsilon}{N+8}\left[ 2a_{\mu}+\left(N+6\right)b_{\mu} \right]    
	\end{split}
	\end{equation}
	The diagonalization of the stability matrix leads to the following
	results for the scaling dimensions:
	\begin{equation}
	\begin{split}
	\lambda_{1}&=3-\frac{6\epsilon}{N+8}+\mathcal{O}(\epsilon^2)\\
	\lambda_{2}&=3+\mathcal{O}(\epsilon^2)    
	\end{split}
	\label{eq_sigma_4-eps}
	\end{equation}

	The eigenvectors of the stability matrix are also interesting because
	they characterize the vector operators associated with each of these
	scaling dimensions. This leads us to introduce the combinations
	\begin{align}
	a'_\mu&=a_\mu-b_\mu \nonumber\\
	b'_\mu&=a_\mu+2 b_\mu
	\end{align}
	which diagonalize the flows:
	\begin{align}
	\partial_{t}a'_{\mu}&=\left(3-\frac{6 \epsilon}{N+8}\right)a'_{\mu}\nonumber\\
	\partial_{t}b'_{\mu}&=3 b'_{\mu}.
	\label{eq:fpfeEps}
	\end{align}
	Rewriting the perturbation hamiltonian, Eq.~(\ref{eq_deltas}), in
	terms of these combinations, we obtain (up to an integration by parts
	which simplifies the $b'_\mu$ term):
	\begin{equation}
	\label{eq_deltas2}
	\int d^dx \Big\{\frac{a'_{\mu}}{6}\phi_i \partial_{\nu}\phi_j \big[\partial_\mu\phi_i\partial_\nu\phi_j-\partial_\nu\phi_i\partial_\mu\phi_j\big]
	+\frac{b'_{\mu}}{24}\phi_i\phi_i\phi_j \partial^2\partial_\mu\phi_j\Big\}.
	\end{equation}
	
	In the Appendix \ref{Ap:redundant}, we show that the second scaling dimension of Eq.~(\ref{eq_sigma_4-eps}) is  $3$ at all orders of
		perturbation theory, as a consequence of a non-renormalization theorem.
		This results from the fact that the associated operator, proportional to $b'_\mu$ in Eq.~(\ref{eq_deltas2}) is redundant.
	We conclude this section by discussing an apparent paradox. In the
	$N=1$ case, there exists only one integrated vector operator. Indeed,
	in this case, the coefficients of $a_\mu$ and $b_\mu$ in the integrand
	of Eq.~(\ref{eq_deltas}) are equal. This seems in conflict with the
	fact that we found two scaling dimensions which are perfectly regular
	in the limit $N\to 1$. However, the term proportional to
	$a'_\mu$ in Eq.~(\ref{eq_deltas2}) vanishes for $N=1$. The scaling
	dimension associated with $a'_\mu$ must therefore be rejected and we
	are left with $\lambda_2$ only, which resolves the paradox.
	
	\subsection{The large-$N$ limit of vector scaling dimensions}
	
\begin{figure}[!ht]
		\begin{center}
			\includegraphics[width=1\textwidth]{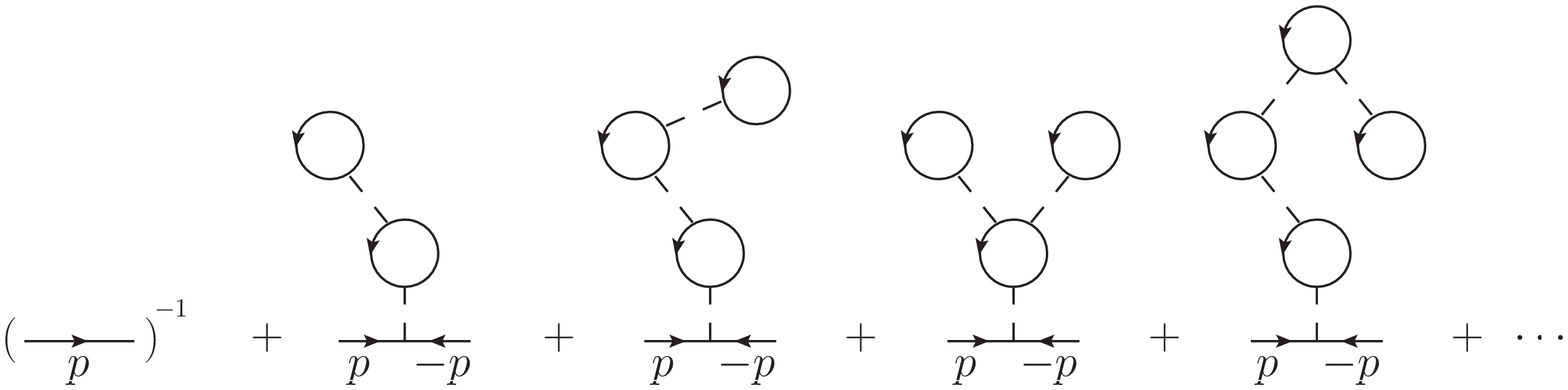}
		\end{center}
		\caption{Leading contribution to $\Gamma^{\left(2\right)}$ in
			a $N^{-1}$ expansion.}
		\label{fig_1}
	\end{figure}

\begin{figure}[!ht]
		\begin{center}
			\includegraphics[width=1\textwidth]{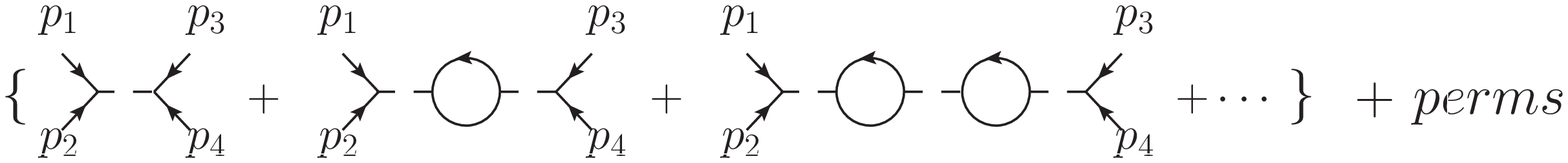}
		\end{center}
		\caption{Leading contribution to $\Gamma^{\left(4\right)}$ in a $N^{-1}$ expansion.}
		\label{fig_3}
	\end{figure}

\begin{figure}[!ht]
		\begin{center}
			\includegraphics[width=1\textwidth]{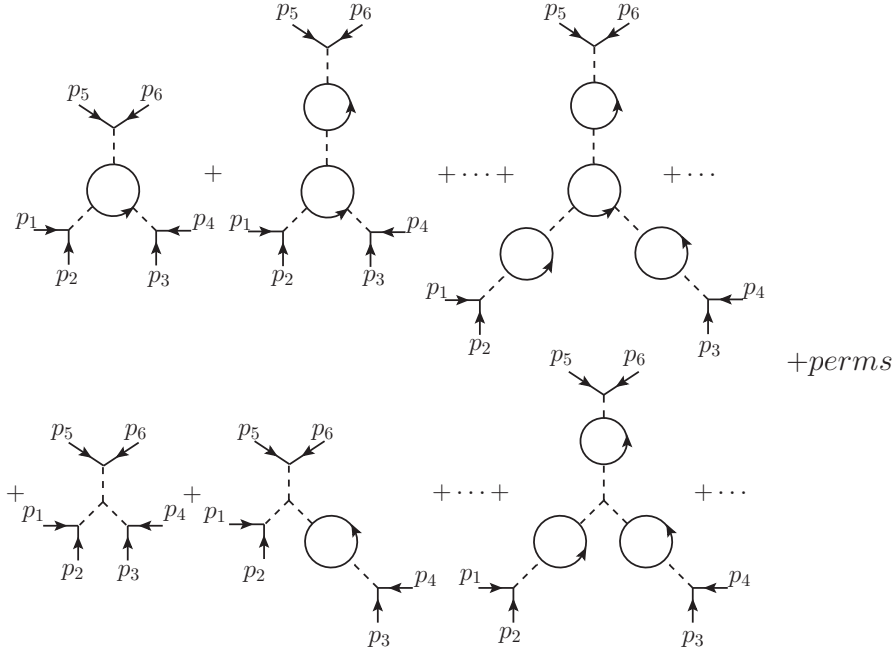}
		\end{center}
		\caption{Leading contribution to $\Gamma^{\left(6\right)}$ in a $N^{-1}$ expansion}
		\label{fig_3.5}
	\end{figure}

	For the large-$N$ limit, we consider the action $S$ perturbed with a generalization of \eqref{eq_deltas} given by:
	\begin{align}
	\label{eq_deltaswithC}
	\delta S= \int d^{d}x \,&\left\{\frac{a_{\mu}}{4}\phi_i \partial^{\mu}\phi_i \partial_\nu\phi_j\partial_\nu\phi_j+\frac{b_{\mu}}{2}  \phi_i\partial_{\nu}\phi_i \partial_{\nu} \phi_j  \partial^{\mu}  \phi_j \right.\nonumber\\
	&+\left.\frac{c_{\mu}}{4}  \phi_i\phi_j\phi_k\partial_{\mu}\phi_i \partial_{\nu} \phi_j \partial_{\nu} \phi_k \right\}
	\end{align}
	As usual, the large-$N$ limit is performed at fixed
	$\hat u=u N$.\footnote{In fact, when considering the full large-$N$
		limit in the presence of couplings $a_\mu$, $b_\mu$ and $c_\mu$ it is necessary
		to also rescale them (see Appendix \ref{Ap:LargeN}). However, this
		rescaling plays no role when considering the flows linearized in
		$a_\mu$, $b_\mu$ and $c_\mu$ [see Eq.~(\ref{eq:feN})]. For simplicity, we ignore this rescaling below.} The diagrams contributing at leading order in the
	large $N$ expansion of correlation functions are well-known (see for example, \cite{Moshe:2003xn}).\footnote{In the case of multicritical fixed points, the large $N$ limit can be more subtle, see \cite{Yabunaka:2018mju}.} These are
	schematically depicted for the two, four and six-point vertices in Figs.~\ref{fig_1}, \ref{fig_3} and~\ref{fig_3.5} respectively. In the last two, it
	is understood that the propagators are effective propagators where all
	cactus diagrams contributing at leading order to $\Gamma^{(2)}$ have
	been re-summed.
	The 4-point (resp. 6-point) interaction  is represented here by two (resp. three) full lines
	connected with dotted lines (the full lines representing the
	Kronecker $\delta$ in vector indices). Note also that we only need to
	work at vanishing external field for determining the scaling dimension
	of the vector operators, which considerably simplifies the
	calculation.
	
	To proceed, we have to compute the 4-point vertex and extract the part
	linear in $a_\mu$, $b_\mu$ and $c_\mu$. As shown in Appendix~\ref{Ap:LargeN}, the 6-point vertex associated with $c_\mu$ gives no contribution to the 4-point vertex function and only diagrams with 4-point vertices contribute, as depicted in Fig.~\ref{fig_3}. Moreover, since we keep only terms linear in $a_\mu$ and $b_\mu$, exactly one vertex of each diagram of Fig.~\ref{fig_3} must be replaced by
	the perturbation ($a_\mu$ or $b_\mu$). A major simplification takes
	place at this point: All diagrams where the $a_\mu$ and $b_\mu$
	couplings appear in a vertex in the middle of a diagram (that is if
	they connect two closed loops) turn out to be zero. Stated otherwise,
	the perturbation can only occur when it is connected to an external leg. This
	result is proven in Appendix~\ref{Ap:LargeN}.
	
	Next we need to compute the part of the 6-point vertex that is just linear in $c_\mu$. There could be contributions coming from $a_\mu$ and $b_\mu$, but these are unimportant for the scaling dimensions because, as mentioned above, there is no  contribution linear in $c_\mu$ to the flows of $a_\mu$ and $b_\mu$, which makes the stability matrix triangular. We denote $X_a$ and $X_b$ the contributions linear in $a_\mu$ and $b_\mu$ to the flow of $c_\mu$.
	
	We consider, again, dimensionless couplings but keep the same symbols as for the dimensionful quantities in
	order to simplify the notations. The resulting flows, which are derived in detail in
	Appendix \ref{Ap:LargeN} are:
	
	\begin{align}
	\partial_{t}\hat{u}&=\left(d-4\right) \hat{u} +\frac{\hat{u}^{2}J}{3}+\frac{\hat{u}^3}{18}L\nonumber\\
	\partial_{t}a_{\mu}&=\left(d-1\right)a_{\mu}+\frac{\hat{u} J}{3}a_{\mu}+\frac{\hat{u}^2}{18}L a_{\mu}\nonumber\\
	\partial_{t}b_{\mu}&=\left(d-1\right)b_{\mu}+\frac{\hat{u} J}{3}b_{\mu}+\frac{\hat{u}^2}{18}L b_{\mu}\nonumber\\
	\partial_{t}c_{\mu}&=\left(2d-3\right)c_{\mu}+X_a a_{\mu}+X_b b_{\mu}+\hat{u} J c_{\mu}+\frac{\hat{u}^2}{6}L c_{\mu}
	\label{eq:feN}
	\end{align}
	\noindent where $J$ is, as before, the dimensionless version of $\int_{q}\dot{R_{k}}\left(q\right)G_{k}^{3}\left(q\right)$. The integral $L$ is
	the dimensionless version of the integral
	$$
	-\int_{q}\dot{R_{k}}\left(q\right)G_{k}^{2}\left(q\right)\times \int_{q'}G_{k}^{3}\left(q'\right)
	$$
	As in the previous section, one can find the fixed-point solution $\hat{u}_*$ and substitute it in the flow for $a_{\mu}$, $b_{\mu}$ and $c_{\mu}$, obtaining
	\begin{align}
	\partial_{t}a_{\mu}&=3 a_{\mu}, \nonumber\\
	\partial_{t}b_{\mu}&=3 b_{\mu}, \nonumber\\
	\partial_{t}c_{\mu}&=X_a^* a_{\mu}+X_b^* b_{\mu}+(9-d) c_{\mu},
	\label{eq:fpfeN}
	\end{align}
	\noindent which implies that  there are two operators with scaling dimension $3+\mathcal O (1/N)$ and one with scaling dimension 9-d. In Eq.~\eqref{eq:fpfeN} $X_a^*$ and $X_b^*$ are the fixed point values. Note that when
	$d=4-\epsilon$, this limit coincides with the large-$N$ limit of the
	$\epsilon$-expansion given in
	Eq~(\ref{eq_sigma_4-eps}). Eq.~\eqref{eq:fpfeN} also implies that
	higher corrections in the $\epsilon$-expansion are all suppressed by
	at least one power of $N^{-1}$ in comparison to the tree-level
	expression. Note that one of these
	eigenvalues is $3$ in all dimensions and for all $N$, due to the
	non-renormaization theorem shown in the Appendix \ref{Ap:redundant}, whereas the
	independence of the other eigenvalue with respect to dimension is
	specific to the large-$N$ limit.
	
	\subsection{Derivative Expansion at order $\mathcal{O}(\partial^3)$}
	
	So far we have considered two limiting cases corresponding to the regimes
	$4-d\ll 1$ and to the large-$N$ limit.  We now implement an
	approximation scheme which is exact in these two limits and which
	remains reasonably accurate for intermediate values of $N$ and $d$.
	In order to do so, we consider the DE approximation of the NPRG at
	order $\mathcal{O}(\partial^3)$.  The DE procedure consists in taking
	an \textit{ansatz} for the effective action $\Gamma_k[\phi]$ in which
	only terms with a finite number of derivatives of the fields
	appear. Equivalently, in Fourier space, it corresponds to expanding
	all proper vertices in power series of the momenta and truncating to a
	finite order.  This approximation is well-suited for studying the
	long-distance properties of the system since higher momentum
	dependence are neglected. In fact, it proved to be a good
	approximation scheme for $\mathbb{Z}_2$ and $O(N)$ models with a very
	good level of precision (see for example,
	\cite{Berges:2000ew,Canet:2003qd,Canet:2002gs,Balog:2019rrg}). The validity of this approximation has been
	discussed in \cite{Blaizot:2005xy,Benitez:2011xx}. It was argued that the NPRG equations have a dressed one-loop  structure where all
	propagators are regularized in the infrared, ensuring the smoothness
	of the vertices as a function of momenta and allowing such an
	expansion. Moreover, the loop diagrams include the derivative of the
	regulating function $\partial_t R_k(q)$ in the numerator. This implies
	that all internal momenta are dominated by the momentum range
	$q \lesssim k$. In consequence an expansion in all momenta (internal
	and external) gives equations that couple only weakly to the regime
	of momenta $p\gg k$. In each model the radius of convergence of the
	expansion in momenta is different but in $O(N)$ models it has been shown
	to be of the order  $q/k\simeq 2-3$ \cite{Balog:2019rrg}. This is consistent
	with the fact that DE shows a rapid apparent convergence at low orders
	for $O(N)$ models.  In fact, the DE has
	been pushed with success to the order $\mathcal{O}(\partial^4)$
	\cite{Canet:2003qd} and $\mathcal{O}(\partial^6)$
	\cite{Balog:2019rrg} for the Ising universality class, giving
	excellent results that improve significantly with the order of the DE.
	
	The procedure then consists in taking the most general terms with the
	symmetries of a given universality class.  In the case of the $O(N)$
	critical regime, we will require invariance under space isometries and
	under the internal $O(N)$ symmetry. To be explicit, in the $O(N)$
	model, the lowest order approximations are: \\ \indent $\bullet$ The
	Local Potential Approximation (LPA) or order $\mathcal{O}(\partial^0)$
	which consist in taking no derivative of the field except a bare,
	unrenormalized, kinetic term
	\begin{equation}
	\Gamma_k=\int d^dx \Big\{ U_k(\rho)+\frac{1}{2}\partial_\mu\phi_i\partial_\mu\phi_i\Big\},
	\end{equation}
	Here, the running effective potential $U_k(\rho)$ is an arbitrary
	function of $\rho=\phi_i\phi_i/2$ whose evolution with $k$ is
	determined by inserting the LPA \textit{ansatz} into the NPRG equation
	(\ref{wettericheq}).
	
	\indent $\bullet$ The $\mathcal{O}(\partial^2)$, which is the
	next-to-leading order, consists in taking all the possible terms
	compatible with the internal symmetries of the system and with at most
	two derivatives. In this case the \textit{ansatz} is
	\begin{equation}
	\Gamma_k=\int  d^dx\Big\{ U_k(\rho)+\frac{1}{2}Z_k(\rho)\partial_\mu\phi_i\partial_\mu\phi_i
	+\frac{1}{4}Y_k(\rho)\partial_\mu\rho\partial_\mu\rho \Big\}.
	\end{equation}
	Again, the functions $U_k(\rho)$, $Z_k(\rho)$ and $Y_k(\rho)$ are
	obtained by inserting the $\mathcal{O}(\partial^2)$ \textit{ansatz}
	into NPRG equations. For $N=1$ the terms in $Z_k(\rho)$ and $Y_k(\rho)$ are equivalent
	and, accordingly, we only include the $Z_k(\rho)$ function.
	
	At higher order in the DE in a rotational invariant scalar model, only even powers of the derivatives appear. However, in the present work, we want to introduce
	terms that have the quantum numbers of a vector while preserving the
	$O(N)$ and translation symmetry. Such terms have necessarily an odd number of
	derivatives and break rotational invariance. Note that there is no $O(N)-$invariant term with a single
	derivative. Indeed, such a term would read
	$$\int d^dx \,  \phi_i\partial_\mu \phi_i K(\rho),$$
	but this is the integral of a total derivative and vanishes. Therefore, we consider here the
	$\mathcal{O}(\partial^3)$ order of the DE that includes all possible
	independent terms with, at most, three derivatives. This is the lowest
	possible DE expansion for our concern. Our \textit{ansatz} reads
	\begin{align}\label{eq:num1}
	\nonumber \Gamma_{k}=&\int  d^{d}x  \left\lbrace U_{k}\left(\rho\right)
	+\frac{1}{2}Z_{k}\left(\rho\right) \partial_\mu\phi_i\partial_\mu\phi_i+\frac{1}{4}Y_{k}\left(\rho\right)\partial_\mu\rho\partial_\mu\rho
	\right. \\
	& \left.+ \frac{1}{4}a_{\mu}\left(\rho\right)\partial_\mu\rho\partial_{\nu}\phi_i\partial_{\nu}\phi_i
	+\frac{1}{2}b_{\mu}\left(\rho\right)\partial_{\mu}\phi_i\partial_{\nu}\phi_i\partial_{\nu}\rho
	+\frac{1}{4}c_{\mu}\left(\rho\right)\partial_{\mu}\rho\partial_{\nu}\rho\partial_{\nu}\rho\right\rbrace.
	\end{align}
	Again, $N=1$ is particular because, in that case, the three structures
	proportional to $a_{\mu}$, $b_{\mu}$ and $c_{\mu}$ are three writings of the same term. As a consequence, in this case,
	we retain only the first one and set $b_{\mu}=c_{\mu}=0$ (on top of seting
	$Y_k=0$ as previously explained). The terms present in $\Gamma_k$
	are of two types. The terms including the functions $U_k$, $Z_k$ and $Y_k$, already present at order $\mathcal O (\partial^2)$, are invariant under rotations and will be called ``isometric-invariant''. The terms including the functions $a_{\mu}$, $b_{\mu}$ and $c_{\mu}$ breaks the rotational invariance.
	
	The calculation proceeds as follows. We first derive the flow equation
	for $U_k$, $Z_k$ and $Y_k$ for vanishing perturbations
	$a_\mu=b_\mu=c_\mu=0$. The flow equation for $U'_k$ is extracted from
	the flow of $\Gamma_k^{(1)}$ in a homogeneous field while the flows of
	$Z_k$ and $Y_k$ are extracted from the flow of the $\mathcal O(p^2)$
	part of $\Gamma_k^{(2)}(p)$ (evaluated again in a homogeneous
	configuration). Observe that there are two independent tensor
	structures in $\Gamma_k^{(2)}(p)$ ($\delta_{ij}$ and $\phi_i\phi_j$),
	which indeed allows us to define the two RG flows of $Z_k$ and $Y_k$. This
	calculation has been done several times in the past (see, for example, \cite{VonGersdorff:2000kp})
	and we verified that we recover the corresponding flow equations and the reported values of the
	critical exponents $\eta$ and $\nu$ for various $N$ at $d=3$
	\cite{VonGersdorff:2000kp,Canet:2002gs}.

	The flow of the perturbations $a_\mu$, $b_\mu$ and $c_\mu$ are then
	extracted from the flow of the $\mathcal O(p^3)$ part of
	$\Gamma_k^{(2)}(p)$ (evaluated again in a homogeneous
	configuration). This can be done because the $\mathcal O(p^3)$ part of
	this vertex (which we note $\Gamma^{(3)}_{p^3}$ below) has three
	independent tensorial structures:
	\begin{align}\nonumber
	\Gamma^{(3)}_{p^3,i_1,i_2,i_3}(p_1,p_2)=\bigg\lbrace&\delta_{i_1 i_2}{\phi_{i_3}}\bigg[-i\frac{a_{\mu}-b_\mu}{2}p_3^{\mu}\big(p_{1}\cdot p_{2}\big)\\
	&\nonumber+i\frac{b_{\mu}}{2}\left[ - p_{1}^{\mu}p_{1}^{2}-
	p_{2}^{\mu}p_{2}^{2}-p_3^{\mu}p_3^{2}\right] \bigg]+\text{2 perms}\bigg\rbrace\\
	&-i \frac{c_\mu}{2}\phi_{i_1}\phi_{i_2}\phi_{i_3}\big[p_1^\mu p_1^2+p_2^\mu p_2^2+p_3^\mu p_3^2\big]
	\end{align}
	where the two permutations are circular permutations of the external
	indices 1, 2 and 3 and where $p_3$ is fixed by momentum conservation.  The resulting flow equations for the six
	functions are treated numerically. All the numerical details are
	considered in Appendix~\ref{Ap:NumParam}.
	
	In order to estimate the error of our prediction of scaling
	dimensions, we proceed in a very conservative way. We consider as
	the central value of our predictions the results obtained in the
	$\mathcal{O}(\partial^3)$ approximation described previously. To evaluate the error bar, we analyze the  poorer approximation where the isometric-invariant sector
	is treated at order LPA (that is, we set in the flow equations
	$Y_k=0$, $Z_k=1$). A first, very pessimistic, estimate of the errors bars
	is to take the double of the difference between these two sets of results.
	This procedure, however, must be slightly improved because it might be that the
	predictions of the two approximation schemes cross accidentally for some value of $N$ and
	$d$. For these exceptional cases, our estimated error bar would vanish,
	which is not reasonable. Accordingly, to cure this problem, we recall that our approximations become exact when $d\to 4$. As a consequence,
	we expect the error to decrease when $d$ grows (at least for $d\ge 2.5$). Thus, we choose the error bars in such a way that it can only grow or remain constant when the dimension is lowered (for $d\ge 2.5$) . This estimate is certainly very pessimistic but we prefer to keep conservative estimates of error bars.
	
	\begin{figure}
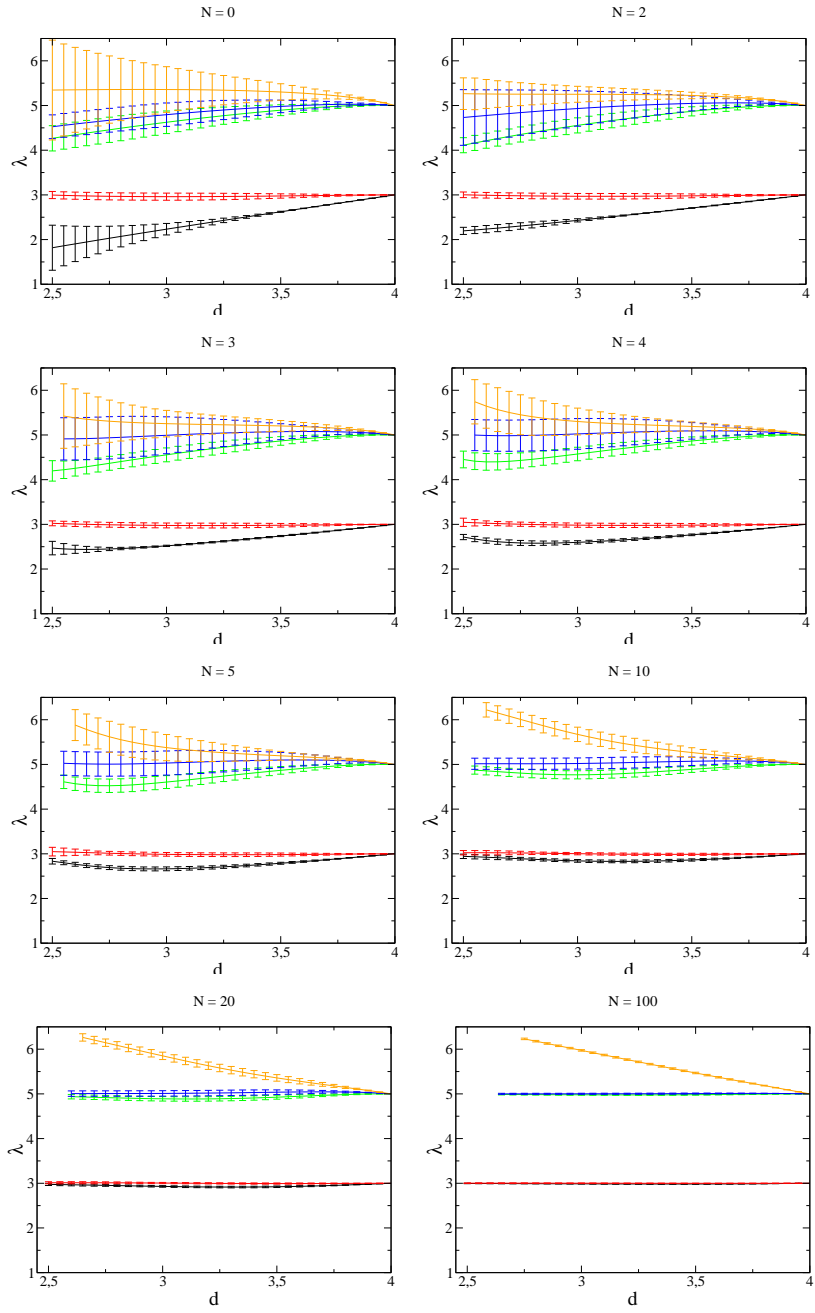

		\centering
		\includegraphics[width=.43\textwidth]{lambdasN0.eps}\hspace{.2cm}
		\includegraphics[width=.43\textwidth]{lambdasN2.eps}
		
		\vspace{.2cm}
		\includegraphics[width=.43\textwidth]{lambdasN3.eps}\hspace{.2cm}
		\includegraphics[width=.43\textwidth]{lambdasN4.eps} 
		
		\vspace{.2cm}
		\includegraphics[width=.43\textwidth]{lambdasN5.eps}\hspace{.2cm}
		\includegraphics[width=.43\textwidth]{lambdasN10.eps}
		
		\vspace{.2cm}
		\includegraphics[width=.43\textwidth]{lambdasN20.eps} \hspace{.2cm}
		\includegraphics[width=.43\textwidth]{lambdasN100.eps}
		\caption{The five smallest scaling dimensions $\lambda$ obtained in $\mathcal O(\partial^3)$ approximation of the NPRG equations are plotted as a function of dimension for various values of $N$. The strategy for evaluating  error bar is explained
			in the text.}\label{fig:SScal1}
	\end{figure}
	We now present the results obtained from the DE at
	$\mathcal{O}(\partial^3)$ order, focusing first on the case $N\neq 1$. We show in Fig.~\ref{fig:SScal1} the five most relevant scaling
	dimensions of vector operators for various $N$ (see Appendix \ref{Ap:NumParam} for the numerical
	details). We restrict to
	five scaling dimensions because the $\mathcal{O}(\partial^3)$
	approximation is unable to properly describe the operators beyond. This can already be understood at the level of perturbation theory, in $d=4-\epsilon$. In order to have control on the operators of dimension $5+\epsilon$, we would need to retain perturbations with 4 fields and 5 derivatives, which are absent of the truncation considered here, see Appendix \ref{Ap:redundant}. Would we include such $\mathcal O(\partial^5)$ terms, we would obtain 5 extra eigenvalues $\sim 5+\mathcal O(\epsilon)$ instead of the 3 shown in Fig.~\ref{fig:SScal1}. For this reason, we expect that the two low-lying scaling dimensions are correctly described, the next 3 are only qualitatively reproduced and we should not consider higher corrections which are probably not under control in this truncation.

	We observe that in
	all cases one eigenvalue is equal to 3, within error bars, in agreement with the exact result given in the appendix \ref{Ap:redundant}. One may wonder about the origin of the (small)
	departure from the exact result. The
	answer is that, at this order, the vertices of the isometric-invariant sector are
	calculated by including the leading order $\mathcal{O}(\partial^0)$ and
	the next-to-leading order contribution $\mathcal{O}(\partial^2)$. This
	is at odds with the flow of the running of the vector coupling which
	is calculated only at leading order $\mathcal{O}(\partial^3)$.  As a consequence there is a (small) mismatch
	between the flows of the potential and the function $a_\mu$.  In fact,
	this difficulty does not take place if the isometric-invariant sector is treated
	only at leading order (LPA). In this case, both the vector and the
	scalar sectors are treated with the same level of accuracy and the
	exact result for $\lambda_2$ is recovered.
	
	The predictions of the $\mathcal{O}(\partial^3)$ approximation are very close to those of the $\epsilon$-expansion [see Eq.~(\ref{eq:fpfeEps})] for the two lowest eigenvalues: the 
	difference between them is, at most, 5\% for $\lambda_1$ and 2\% for $\lambda_2$,
	for all values of $N$ for $d\ge 3$ (not shown in Fig.~\ref{fig:SScal1}). From Fig.~\ref{fig:SScal1} it is also clear
	that our $\mathcal{O}(\partial^3)$ solution is compatible with the results for the large $N$
	limit and the $\epsilon$-expansion from Eq.~(\ref{eq:fpfeN}) and
	Eq.~(\ref{eq:fpfeEps}) in their respective domains of validity.  In all cases, the
	leading exponent is above $2$ (even for $N=0$). By using the
	sufficient condition discussed in Sect.~\ref{Sec.II}, we therefore conclude that
	conformal invariance is indeed realized at the critical point of
	$O(N)$ models, for all the values of $N\neq 1$ and $d$ that we could
	consider.

\begin{figure}
		\centering
		\includegraphics[width=7.4cm]{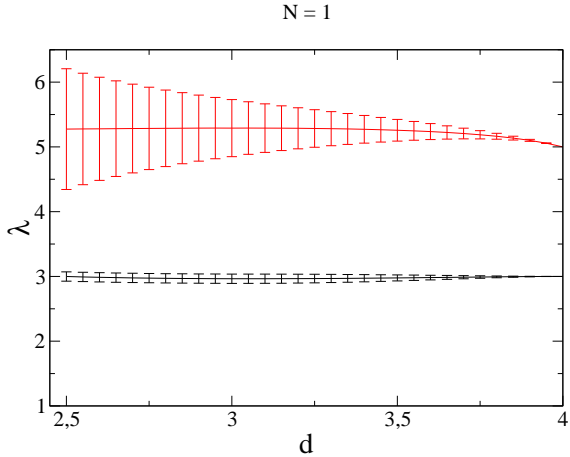} 
		\caption{Scaling dimensions $\lambda_{2}$ and $\lambda_{3}$ for $N=1$ as a function of the space dimension $d\in [2.5,4]$.
			The error bar estimates are explained
			in the text.}\label{fig:SScal1N1}
	\end{figure}

	The case $N=1$ is particular because, as
	already explained in Sect.~\ref{sec_eps}, some eigenvectors do not exist. As a consequence, there is just one eigenvalue which behaves as $3+\mathcal O(\epsilon^2)$ and one which behaves as $5+\mathcal O(\epsilon)$ (we would have three extra eigenvalues $\sim 5+\mathcal O(\epsilon)$ for a richer truncation including terms with 5 derivatives in the perturbation). As discussed previously for generic $N$, we only expect a qualitative description of this eigenvalue.\footnote{The operator with 6
		powers of the field and 3 derivatives has scaling dimension $5+4\epsilon/3 +O(\epsilon^2)$. The operators with 4 powers of the field and 5 derivatives
		have scaling dimension $5 +O(\epsilon^2)$, $5 − 4\epsilon/9 +O(\epsilon^2)$ and $5 − 2 \epsilon/3 +O(\epsilon^2)$.} The two lowest eigenvalues are shown in 
	Fig.~\ref{fig:SScal1N1}. Again, the value -1 can be unambiguously rejected.

	\section{Inequalities for $O(N)$ models}
	
	\subsection{Review of the proof for the Ising model}\label{Sec.III}
	
	The results of the previous sections give strong indications that the
	scaling dimension of integrated vector operators are always larger than
	$-1$.  The problem with such a reasoning is that one could, in
	principle, doubt about the meaning of the theory in arbitrary real
	dimensions $d$.  Even if it clearly makes sense to all orders of
	perturbation theory around $d=4$, the non-perturative meaning of this
	objects is not evident (see, however,
	\cite{Berges:2000ew,Codello:2014yfa,El-Showk:2013nia}). In the same way,
	we could distrust the quality of the $\epsilon-$expansion, in $d=3$, or the
	quality of $\mathcal O (\partial^3)$ or large $N$ approximations. Accordingly,
	in order to have a more rigorous proof for models in physical
	dimensions below the upper critical dimension, it is convenient to
	have a more robust control of scaling dimensions of vector
	operators. With this objective in mind, a lower bound for the scaling
	dimension of the vector operators which are scalar under internal symmetries was found for the Ising
	universality class \cite{delamotte2016scale}. We review this proof in
	this section and add some material that was not present in the
	original proof.
	
	In this section, and contrarily to what we did so far, we consider
	local operators. Instead of studying the RG flow around a fixed point, we will extract the scaling dimension of the vector operator by considering the power-law decay of the correlation between two vector operators as a function of the distance. For most operators, scale invariance
	implies that the two-point connected correlation function behaves at
	criticality as:
	\begin{equation}
	\label{powerlawvect}
	\left\langle\mathcal{V}_{\mu}(x)\mathcal{V}_{\mu}(y)\right\rangle_{c} \sim \frac{1}{|x-y|^{2D_\mathcal{V}}}
	\end{equation}
	for large enough $|x-y|$. If one proves that for any
	$\mathcal{V}_{\mu}(x)$, one has $D_\mathcal{V}>d-1$, any integrated
	vector operator of the form
	$\int d^dx \mathcal{V}_{\mu}(x)$ have scaling dimension
	$D_\mathcal{V}-d>-1$. 
	
	It is important to stress that this bound is
	very pessimistic because we do not discard total derivatives (that
	would not contribute to the integrated operator for reasonable
	boundary conditions). In this sense, we give a bound for a family of
	operators which is larger than the one we are interested in (those
	which are not total derivatives). In particular, the bound is not
	satisfied in $d=4$ where there is a total derivative local vector
	operator, namely $\partial_\mu (\varphi^2)(x)$, that has dimension exactly
	$d-1=3$. However, we will see that even this leading local operator
	has dimension strictly larger than $d-1$ in any dimension
	below four. More generally, we will prove that any local vector
	operator even in $\varphi$ has scaling dimension strictly larger than $d-1$ for any $d<4$.

	A complication appears in the proof because some  operators may have correlation function which do not behave as a power-law as in Eq.~\eqref{powerlawvect}, but as contact operators (that is, they are $\delta$ correlated; see for example \cite{Nakayama:2019mpz}). This for instance occurs when an operator is proportional to the equations of motion. To circumvent this issue, we can use our freedom of rewriting the integrated operator in terms of different densities (which differ by total derivatives) to ensure that the correlation have a power-law decay. To be more explicit, let us consider
	as an example the vector operator 
	\begin{align}
	\mathcal V_\mu&= \frac{\delta S}{\delta \phi(x)} \partial_\mu \partial^2\phi\\
	&=(-\partial^2 \phi+r\phi+\frac u6\phi^3 )\partial_\mu \partial^2\phi\label{eq_toy}
	\end{align}
	with the action $S$ defined by Eq.~(\ref{action}), rewritten for the Ising universality class. This operator has the right symmetries but being proportional to the equation of motion, the correlation functions of such operators are delta correlated and the bounds derived below on the correlation functions at long distance are not useful. A way out consists in considering instead the operator $\frac u2 \phi\partial_\mu\phi(\partial_\nu \phi)^2$, which differs from the one introduced in Eq.~(\ref{eq_toy}) by total derivatives, but which has a power-law decay at long distances (this can be checked already at tree-level). In what follows, we assume that an integrated vector operator can always be rewritten as the integral of a density whose correlation functions have a power-law behavior at long distances.

	In order to prove the inequality, we start with the Ginzburg-Landau
	model in a cubic lattice with lattice spacing $a$ and a $\varphi^4$ interaction. This model is in
	the Ising universality class. It follows from Griffiths and Lebowitz
	inequalities
	\cite{griffiths1967correlationsI,kelly1968general,lebowitz1974ghs}
	that, at zero external magnetic field and for any temperature
	$T\ge T_c$:
	\begin{equation}
	\label{boundcorrelIsing}
	0\leq \left\langle\varphi^n(x)\varphi^m(y)\right\rangle -\left\langle\varphi^n(x)\right\rangle\left\langle\varphi^m(y)\right\rangle\leq\left\{
	\begin{array}{ll}
	C(n,m) G(x-y)\hspace{.5cm}&\mathrm{if\, n\, and\,m\,odd}\\
	\hat{C}(n,m) G^2(x-y)&\mathrm{if\, n\, and\,m\,even}
	\end{array}
	\right.
	\end{equation}
	where $G(x-y)=\langle\varphi(x)\varphi(y)\rangle$, $n$ and $m$ are integers and $C(n,m)$ and $\hat{C}(n,m)$ are constants. This inequality was first proven in \cite{sokal1981more} and was independently rederived
	in \cite{delamotte2016scale}.
	
	By using scale invariance at the critical point, one knows that
	$\left\langle\varphi^n(x)\varphi^m(y)\right\rangle -\left\langle\varphi^n(x)\right\rangle\left\langle\varphi^m(y)\right\rangle$ behaves as a power-law
	of $|x-y|$. Moreover at the critical point, $G(x-y)\propto |x-y|^{-(d-2+\eta)}$ where $\eta$ is the anomalous dimension of the operator $\varphi$.
	This imply that, for $m$ and $n$ even and at the critical point:
	\begin{equation}
	\Big|\left\langle\partial_\mu[\varphi^n(x)]\partial_\nu[\varphi^m(y)]\right\rangle \Big| \leq \frac{\tilde{C}(n,m,\mu,\nu)}{|x-y|^{2(d-1+\eta)}}
	\end{equation}
	where derivatives are a shortcut notation for appropriate finite difference
	expressions defined in the lattice.\footnote{As it will be explained
		below it is important to choose centered finite differences in order
		to ensure that the vector operators are odd under parity.} As a
	consequence, for all operators
	$\mathcal{V}_{\mu}(x)=\partial_\mu(\varphi^n(x))$ with $n$ even, one
	has $D_\mathcal{V}\ge d-1+\eta$.
	
	So far, all the elements of the proofs are completely rigorous but the
	next step requires to make an assumption that we describe now.  We consider two
	different discretizations $\mathcal{O}^{(a)}_1(x)$ and
	$\mathcal{O}^{(b)}_1(x)$ of a given operator in the continuum. We,
	moreover, ask these discretized operators to have the same
	transformation rules as the operator in the continuum under the
	group of internal symmetries and under mirror images about the lattice
	planes. Then, for arbitrary operators
	$\mathcal{O}_2(x), \mathcal{O}_3(x), \dots$, we assume that
	\begin{equation}
	\label{assumption}
	\left\langle \mathcal{O}^{(a)}_1(x_1)\mathcal{O}_2(x_2)
	\dots \mathcal{O}_n(x_n)\right\rangle
	=Z_{a,b} \left\langle \mathcal{O}^{(b)}_1(x_1)\mathcal{O}_2(x_2)
	\dots \mathcal{O}_n(x_n)\right\rangle
	\end{equation}
	when the various points $x_1, x_2, \dots,x_n$ are far apart (as
	compared to the lattice spacing $a$). It is not difficult to prove that
	this assumption is true to all orders in perturbation theory. We will
	also discuss below its validity beyond perturbation theory. Before
	doing so, let us discuss its consequences for the end of the proof of
	the validity of conformal invariance.
	
	The previous assumption implies, in particular, that for any set of vectors $e_i$ belonging to the lattice, any operator with the form
	\begin{equation}
	\mathcal{W}_{\mu}(x)=\frac{1}{2}(\partial_{\mu}\varphi)(x)\sum_{s=\pm 1}\prod_{i=1}^{n-1}\varphi(x+se_i)
	\label{eq.formVecOp}
	\end{equation}
	have the same large distance behaviour (with respect to the
	lattice spacing) as the total derivative operator
	$\varphi^{n-1}(x)\partial_{\mu}\varphi(x)$ (modulo a multiplicative
	factor) because both of them are discretizations of the same continuum
	operator. One deduces that also for the operators of type
	$\mathcal{W}_{\mu}(x)$ which are even in $\varphi$ one has
	$D_\mathcal{W}\ge d-1+\eta$. Now, {\it any} vector operator even in
	$\varphi$ can be discretized as a linear combination of operators of
	type $\mathcal{W}_{\mu}(x)$. Triangle inequality then implies that for
	{\it any} vector operator even in $\varphi$ one has
	$D_\mathcal{V}\ge d-1+\eta$. Since $\eta$ has been proven to be
	strictly positive in an interacting theory whose minkowskian extension
	is unitary \cite{ZinnJustin:2002ru,Pohlmeyer:1969av}, one deduces that
	any local vector operator even in $\varphi$ satisfies
	$D_\mathcal{V}> d-1$ which concludes our proof of conformal invariance in the
	critical regime of the Ising universality class.
	
	Let us now analyze in detail the validity of the assumption
	(\ref{assumption}). First of all, the $\varphi^4$ model is
	super-renormalizable for $d<4$. The lattice can be seen as an
	ultraviolet regularization and the various discretizations of an
	operator can be seen as proper regularization of a given continuum
	operator.  It is important to keep in mind that an operator mixes
	with all operators with lower dimension and the same quantum
	numbers. With a lattice regularization, only quantum numbers preserved
	by the lattice are easy to keep under control. For example, the notion
	of  ``vector operator'' is somewhat ambiguous on the lattice because it
	refers to the properties of a given operator under the symmetries of the continuum
	space. Nonetheless, a subset of space symmetries are preserved by the
	lattice and we can classify operators by studying their transformation
	under these symmetries. In particular, if we work with a cubic lattice,
	we can use parity and require that a discretized operator which is a
	regularization of a vector operator must be odd under parity.
	
	As an example, the lattice finite difference
	$\varphi(x)\big(\varphi(x+\hat\mu a)-\varphi(x-\hat\mu a)\big)$ is odd
	under parity and can be considered as a lattice regularization of a
	vector operator. On the contrary,
	$\varphi(x)\big(\varphi(x)-\varphi(x+\hat\mu a)\big)$ is not an odd
	operator. As such, it mixes both with scalar operators and vectors
	operators, even if its na\"ive continuum limit is proportional to
	$\varphi(x)\partial_\mu \varphi(x)$. As pointed out before, it is
	necessary then to restrict the finite differences
	considered for regularization derivatives to be centered in order to
	preserve the proper behaviour under parity.\footnote{One can add
		further requirement, such as other cubic symmetries but we will not
		need them below.} Having preserved the parity of the operators, it is
	clear that operators with an odd number of derivatives can not mix
	with operators with an even number of derivatives. Below we always
	assume that centered differences are used in order to avoid
	difficulties with parity.
	
	As explained before, in the neighbourhood of a fixed point of the RG,
	one can construct a basis of operators which are eigenoperators of
	scale transformations. Accordingly, in the infrared regime of a
	scale-invariant theory, any operator can be decomposed on a basis of
	operators which have the same quantum numbers:
	\begin{equation}
	\mathcal{O}^{(a)}(x)=\sum_i Z_a^{(i)} \mathcal{O}^{(i)}(x)
	\end{equation}
	where the operators $\mathcal{O}^{(i)}(x)$ are renormalized
	operators in the sense that their correlation functions have a limit
	when the lattice spacing $a$ goes to zero at fixed values of the
	distance of the various operators appearing in correlation
	functions. Moreover, the operators $\mathcal{O}^{(i)}(x)$ are chosen
	to be eigenoperators of scale transformations.  Now, two operators
	$\mathcal{O}^{(a)}(x)$ and $\mathcal{O}^{(b)}(x)$ having the same
	quantum numbers can be expanded with the same set of renormalized
	operators $\mathcal{O}^{(i)}(x)$. For example,
	$1$, $\phi^2(x)$, $\phi^4(x)$ and $\phi(x)[\phi(x+a\hat \mu)+\phi(x-a\hat \mu)]$ have the same quantum numbers and they can therefore be decomposed on the same basis. This implies that, as is well-known, the long distance behavior of the correlation functions of $\phi^2(x)$ and $\phi(x)[\phi(x+a\hat \mu)+\phi(x-a\hat \mu)]$ are dominated by the mixing with the identity.

	Let us assume without loss of generality that the most relevant
	operator in this list is $\mathcal{O}^{(1)}(x)$. Accordingly, at large
	distances, $\mathcal{O}^{(a)}(x)$ behaves as
	$Z_a^{(1)} \mathcal{O}^{(1)}(x)$ and $\mathcal{O}^{(b)}(x)$ as
	$Z_b^{(1)} \mathcal{O}^{(1)}(x)$. In consequence, as long as the factors
	$Z_a^{(1)}$ and $Z_b^{(1)}$ are not zero, the two operators
	$\mathcal{O}^{(a)}(x)$ and $\mathcal{O}^{(b)}(x)$ have the same large
	distance behaviour.
	
	Given that $\mathcal{O}^{(a)}(x)$ and $\mathcal{O}^{(b)}(x)$ are two
	regularizations of the same ``continuum'' operator, if the
	renormalization factor $Z_a^{(1)}$ is non-zero, typically the
	renormalization factor $Z_b^{(1)}$ is also non-zero. The Feynman
	diagrams associated with both operators are essentially the same and
	only combinatoric coincidences can make one of these coefficients to
	be zero at leading order of perturbation theory. Accordingly, barring
	these exceptions, in the realm of perturbation theory both operators
	have the same critical exponents.
	
	We could not prove that the assumption is valid beyond perturbation
	theory. However, in this regime, the assumption is equivalent to
	the assumption that, adding an even number of derivatives
	(or, more precisely, centered finite differences) to a given operator leads to an operator
	which is equally or less relevant, but not more relevant.  This is
	intimately related to the existence of a continuum limit. Indeed, if
	adding more and more derivatives would give rise to an operator which
	is {\it more relevant} the notion of universality would be completely
	lost. In particular, all Monte-Carlo simulations would be under
	suspicion. If higher derivative operators could be more relevant, two
	different discretization of a given operator would have different
	scaling properties.  Finally, we mention that, for scalar theories in
	$d=2$ and $d=3$ the continuum limit does exist (see, for example, \cite{Fernandez:1992jh}).  Now, even if typical
	configurations in a functional integral are not differentiable, in
	order for this continuum limit to exist, some level of regularity of
	the configurations is necessary. Accordingly, we should expect that adding an
	even number of derivatives should lead to operators that are, at most,
	equally relevant, but not more relevant.
	
	We stress that the previous discussion only applies to adding an {\it
		even} number of derivatives. Indeed, adding a single derivative can lead to
	an operator that is more relevant. This is for instance the case of
	the operator $\varphi\partial_\mu \varphi$. If we add a single extra
	derivative we can generate the operator
	$\partial_\mu\varphi\partial_\mu\varphi$ which has the same quantum
	numbers as $\varphi^2$ and that, in fact, behaves as $\varphi^2$ at
	large distances (which is more relevant than
	$\varphi\partial_\mu \varphi$). This, however, does not play any role
	in our proof of conformal invariance as long as we use centered
	derivatives. Indeed, we only need to assume that operators with an
	odd number of derivatives are not more relevant that operators with
	a single derivative.

	\subsection{Extension of the proof for some $O(N)$ models}\label{Sec.IV}
	\label{extendingproof}

	There is an extensive bibliography on correlation inequalities which were used in many cases to prove some properties of statistical
	systems \cite{griffiths1967correlationsI,griffiths1967correlationsII,griffiths1967correlationsIII,sokal1981more,krinsky1974upper,griffiths1969rigorous}.
	We use here a generalization of the inequality (\ref{boundcorrelIsing}) valid for (at least some) $O(N)$ models. 
	This can be achieved by making use of a generalization of the Griffiths and Lebowitz
	inequalities \cite{bricmont1977gaussian,dunlop1979zeros,dunlop1976correlation,monroe1979correlation,kunz1975correlation,sokal1982mean} to
	$O(N)$ models when $N=2, 3,$ or $4$. To be explicit, it is shown in Appendix \ref{inequalities} that, at vanishing external magnetic field and for~$T\geq T_{c}$,
	\begin{align}
	\nonumber \Big| &\langle \varphi_{i_1}\left(x\right) \cdots \varphi_{i_m}   \left. \left(x\right)\varphi_{j_1}\left(y\right)\cdots  \varphi_{j_n}\left(y\right)\right\rangle  \\ & - \left\langle \varphi_{i_1}\left(x\right)\cdots \varphi_{i_m}\left(x\right)\right\rangle \left\langle\varphi_{j_1}\left(y\right)\cdots \varphi_{j_n}\left(y\right)\right\rangle \Big| \leq \left\lbrace
	\begin{matrix}
	C(n,m) G\left(x-y\right)\textit{n,m}\text{ odd}\\ \hat{C}(n,m) G^{2}\left(x-y\right)\textit{n,m}\text{ even}
	\end{matrix}\right.
	\label{eq:sufCond}
	\end{align}
	where $C(n,m)$, $\hat{C}(n,m)$ are constants and $G(x-y)$ is defined through
	\begin{equation}
	\left\langle \varphi_{i}\left(x\right)\varphi_{j}\left(y\right)\right\rangle=\delta_{ij}G(x-y). 
	\end{equation}
	The inequality (\ref{eq:sufCond}) generalizes for $N=2,3$ and $4$ the inequality (\ref{boundcorrelIsing}) valid for $N=1$.
	
	As for the Ising universality class, we can then use scale invariance
	at the critical point to conclude  that
	$| \left\langle \varphi_{i_1}\left(x\right) \cdots \varphi_{i_m}
	\right. \left. \left(x\right)\varphi_{j_1}\left(y\right)\cdots
	\varphi_{j_n}\left(y\right)\right\rangle - \left\langle
	\varphi_{i_1}\left(x\right)\cdots
	\varphi_{i_m}\left(x\right)\right\rangle
	\left\langle\varphi_{j_1}\left(y\right)\cdots
	\varphi_{j_n}\left(y\right)\right\rangle|$ behaves as a power of
	$|x-y|$ when $|x-y|$ is much larger than the lattice
	spacing. Moreover, since
	$G\left(x-y\right)\propto |x-y|^{-(d-2+\eta)}$ at the critical point,
	we deduce that, for any even $n$ and $m$,
	\begin{equation}
	\label{ineqon}
	\Big|\left\langle \partial_\mu(\varphi_{i_1}\left(x\right) \cdots \varphi_{i_m}\left(x\right)) \right. \left. 
	\partial_\nu(\varphi_{j_1}\left(y\right)\cdots  \varphi_{j_n}\left(y\right))\right\rangle\Big| \leq \frac{\tilde{C}(\mu,\nu)}{|x-y|^{2(d-1+\eta)}}.
	\end{equation}

	An important difference between the Ising universality class and the
	$O(N)$ vector model appears now. Indeed, when a single scalar field is
	present, all operators of type (\ref{eq.formVecOp}) behave in the
	continuum limit as total derivatives. When two or more scalars are
	present this is no longer true. For example, the operator
	$\varphi_1(x)\partial_\mu(\varphi_2(x))$ is {\it not} a total
	derivative. As a consequence, the inequality (\ref{ineqon}) does not
	apply to this operator. More generally, for $O(N)$ models there are many local vector
	operators, even in the number of fields, for which
	the inequality (\ref{ineqon}) can not be applied. We recall however
	that the sufficient condition under which scale invariance implies conformal invariance relies on the scaling dimension of vector operators which are scalars under the $O(N)$ group. Consequently, we are only interested in operators where all internal indices are
	contracted. In that case any operator of the form
	\begin{equation}
	\mathcal{W}_{\mu}(x)=\frac{1}{2}(\partial_{\mu}\varphi_j)(x)\sum_{s=\pm 1}\varphi_{j}(x+s e_0)
	\prod_{i=1}^{m-1}\varphi_{k_i}(x+se_i)\varphi_{k_i}(x+se'_i)
	\label{eq.formVecOpon}
	\end{equation}
	has the same na\"{i}ve continuum limit that a total
	derivative [which generalizes
	Eq.~\eqref{eq.formVecOp}]. We now assume, as done in the Ising universality class (see discussion on the point in the previous section),
	that operators having the same continuum
	limit have the same scaling dimension. We deduce, as in the Ising case, that
	$D_\mathcal{W}\ge d-1+\eta$. Now, any $O(N)$ invariant local operator
	$\mathcal{V}$ can be regularized by a linear combination of
	operators of type \eqref{eq.formVecOpon}. As for the Ising universality class, it is important to avoid contact operators by choosing appropriately the local vector operator $\mathcal{V}_{\mu}(x)$. By invoking again the
	triangle inequality we deduce that for any such operator,
	$D_\mathcal{V}\ge d-1+\eta$ (as long as $N=2, 3,$ or $4$). Again, we
	recall that $\eta$ has been proven to be strictly positive in an
	interacting theory whose Minkowskian extension is unitary
	\cite{ZinnJustin:2002ru,Pohlmeyer:1969av}, and, accordingly, any local
	$O(N)$ invariant vector operator has $D_\mathcal{V}> d-1$, which
	concludes our proof.
	
	It is important to observe that the previous reasoning relies on imposing that candidates for virial current must be invariant under the
	full $O(N)$ group. This ensures that the only invariant tensor is the Kronecker delta. This requirement is fully justified because, as shown
	before, the functional $\Sigma_{\mu}[\phi]$ must be invariant under the full group of internal symmetries of the considered universality class.
	In fact, if one consider more general vectors, as for example, if we relax the constraint to the $SO(N)$ group only, the completely
	anti-symmetric tensor with $N$ entries is also invariant. In this
	case, the last step of our proof (following
	Eq.~(\ref{eq.formVecOpon})) is not valid anymore.  This observation is
	at the heart of the criticism raised in in the recent preprint
	\cite{Meneses:2018xpu} where it was observed that, for $N=2$, one can
	construct the conserved current
	$J_\mu=\varphi_1\partial_\mu\varphi_2-\varphi_2\partial_\mu\varphi_1$
	which is $SO(2)$ invariant and has scaling dimension exactly equal to
	$d-1$. At first sight this would be a counter-example of the present
	proof. However, the current $J_\mu$ is not invariant under the mirror
	symmetry $\varphi_1\to -\varphi_1$ which is an element of the $O(2)$
	group but not of the $SO(2)$ group.
	
	\section{Conclusions} \label{Sec.VI}
	
	In this article, the presence of conformal invariance in critical $O(N)$ models has been considered. For any translational and rotational invariant model, two sufficient conditions for the validity of conformal invariance in presence of scale invariance have been reviewed \cite{delamotte2016scale,polchinski1988scale}. In general these two conditions rely on different hypotheses but for short-range interactions (as the usual $O(N)$ model) they are actually equivalent. Both conditions require that there exist no integrated vector operator which have the same internal (linearly-realized) symmetries as the Hamiltonian and with scaling dimension -1.
	
	With this sufficient condition, the problem reduces to estimating the scaling dimensions of integrated vector operators. For $d\ge 4$, where the scaling dimensions can be calculated exactly (because the theory is controlled by the Gaussian fixed point), it is easy to show
	(see~\cite{delamotte2016scale}) that all integrated vector operators have dimensions much higher than $-1$ (the smallest critical dimensions of an integrated vector operator in $d=4$ has dimension 3). When the space dimension is lowered, as long as we can trust the $\epsilon-$expansion, one can expect critical dimensions to vary moderately. As a consequence it would be surprizing that  $\epsilon$ corrections would change the exponents from 3 to -1 when $d$ goes from $d=4$ to $d=3$. Moreover, as long as the spectrum of critical dimensions is discrete and vary smoothly with $d$ (as usually assumed), Ward-identities for conformal invariance would remain valid for any dimension, except in the space dimension where the critical dimension of a vector operator crosses the value $-1$. Even in that case, one would expect conformal invariance to remain true \cite{delamotte2016scale}, as long as correlation functions depends smoothly with $d$. Indeed, even if a vector operator has critical dimension $-1$ exactly in $d=3$, for any $d>3$ conformal invariance would be valid and, as a consequence, also in $d=3$ by continuity.
	
	These arguments are in favour of the validity of conformal invariance for most critical models in any dimension, but stay at a low level of rigour. As such, a more detailed control of the various reasonable expectations is welcome and, if possible, more rigorous control of the possible values of critical dimensions of vector operators in any dimension. 
	
	In this article, in order to obtain more convincing arguments in favor of conformal invariance, we have computed the lowest scaling dimensions of vector eigenoperators  within three approximation schemes: the $\epsilon-$expansion (at order $\epsilon$), the large $N$ limit and the Derivative Expansion of the NPRG at order $\mathcal{O}(\partial^3)$. The results are in line with the standard expectations: the scaling dimensions vary smoothly with $d$, the spectrum is discrete and the variation of scaling dimensions with $d$ is moderate. The estimates of scaling dimensions obtained from the three approximation scheme are  compatible. For  $d\ge 2.5$ and  $N\ge 1$ the scaling dimension of the lower integrated vector operator is unambiguously larger than 2, which  implies conformal invariance. The extension of $O(N)$ models to the limit $N\to 0$ (relevant for self-avoiding polymer chains \cite{degennes72}) is also considered and, again in this case, conformal invariance is obtained for any $d\ge 2.5$.  In the case $N=1$ of the Ising universality class, our estimates coincides (within error bars) with Monte-Carlo estimates \cite{Meneses:2018xpu} which also exclude the value -1.
	
	In the second part of the article, we have  generalized a proof performed previously in the Ising universality class \cite{delamotte2016scale} to the $O(N)$ models for $N=2,3$ and $4$. For those values of $N$ rigorous
	bounds for correlation functions are known \cite{bricmont1977gaussian,dunlop1979zeros,dunlop1976correlation,monroe1979correlation,kunz1975correlation,sokal1982mean} that generalize the standard Griffiths \cite{griffiths1967correlationsI,griffiths1967correlationsII,griffiths1967correlationsIII,kelly1968general,griffiths1970concavity} and Lebowitz \cite{lebowitz1974ghs} inequalities (valid for Ising universality class). By employing those inequalities, we prove that, under mild assumptions (which concern the continuum limit of a model on a lattice) the scaling dimension of any integrated vector operator is strictly larger than $-1$
	for any $d<4$, implying conformal invariance. 
	
	For the future, some aspects of the present analysis can be extended to other models. For example, the perturbative analysis can be exteded to more involved models such as the model with cubic symmetry. We are planing also to employ the expansion in $d-2$ for $O(N)$ models (with $N>2$). Concerning the proofs based on correlation inequalities, the present analysis may by generalized to self-avoiding polymer chains. In that case some correlation inequalities are known too and we plan to study if they can be employed in order to generalize our results.
	
	\begin{acknowledgements}
		The authors thank B. Delamotte, G. Tarjus,  T. Morris and Y. Nakayama  for fruitful discussions. The authors acknowledge financial support from the ECOS-Sud France-Uruguay program U11E01. N. W. and D. P. thanks PEDECIBA (Programa de desarrollo de las Ciencias B\'asicas,
		Uruguay) and acknowledges funding through grant from the Comisi\'on Sectorial de Investigaci\'on Cient\'ifica de la Universidad de la Rep\'ublica, Project I+D 2016 (cod 412).
	\end{acknowledgements}
	
	\begin{appendix}
		
		\section{Redundant operators}
		\label{Ap:redundant}
		
		This appendix is devoted to a discussion of redundant vector operators, with an emphasis on conformal invariance. Redundant operators are of the form $(\delta S/\delta \phi(x)) \mathcal O(x)$. They are sometimes considered to be physically uninteresting. Indeed, they typically have short-range correlation functions.\footnote{This implies that it is not possible to define their scaling dimension by looking at the power-law behavior of correlation functions at long distances. It is however possible to define a {\it bona fide} scaling dimension by a stability analysis of the renormalization-group flow around the fixed point. We stress that the constraint given in the introduction on the scaling dimension for the operator $\Sigma_\mu$ corresponds to this latter definition.} 
		
		\subsection{Redundant operators and breaking of conformal invariance}
		
		The aim of this section is to explain why redundant operators can be ignored as possible breakings of conformal invariance. On general grounds, a redundant operator, with short-range correlations can be responsible for the breaking of Ward identities of a typical symmetry. The mere existence of such an operator would have strong physical consequences because correlation functions for other fields would not display the corresponding invariance. We illustrate this in a very simple situation. Consider a model with two scalar fields $\phi_1$ and $\phi_2$ whose dynamics is given by a general action $S$ which needs not be O(2)-symmetric. If we perform, in the path integral of the partition function, a change of variable $\phi_i\to\phi_i+\theta \epsilon_{ij}\phi_j$ (here $\epsilon_{ij}$ is the bidimensional Levi-Civita tensor and $\theta$ an infinitesimal angle) which corresponds to an infinitesimal rotation in internal space, we obtain:
		\begin{equation}
		\label{eq_o2}
		\int d^dx\left(\epsilon_{ij}J_i\frac{\delta W}{\delta J_j}\right)=\int d^dx\left\langle \epsilon_{ij}\phi_i \frac{\delta S}{\delta \phi_j}\right\rangle.
		\end{equation}
		The brackets in the right-hand-side represent an average over the fields with the Boltzmann distribution in presence of sources $J_i$ for the fields $\phi_i$. Of course, if the action is O(2) symmetric, we recover the Ward identity for rotation in internal space. However, for a generic action $S$, the right-hand-side does {\it not} vanish and the O(2) Ward Identity is not satisfied. Now, what is of interest for us here is that the right-hand-side of the previous equation is the average of a {\it redundant} operator. The operator $\epsilon_{ij}\phi_i \frac{\delta S}{\delta \phi_j}$ appearing in the right-hand-side of Eq.~(\ref{eq_o2}), which has only contact terms in its correlation functions, is physically important because it induces a breaking of O(2) invariance, at the level of Ward identities.
		
		To make an analogy with the strategy followed in this article to study conformal invariance, suppose we want to prove that a model is invariant under O(2) by searching for putative operators that could appear in the right-hand-side of Eq.~(\ref{eq_o2}). Suppose that we can discard the existence of such operators which are not redundant but that we have no control on redundant ones. Then, the previous example shows that we have no way to conclude on the O(2) invariance of the theory. If, instead, we can discard both non-redundant and redundant operators, then we conclude that the theory is indeed invariant.\footnote{It is often stated in the literature that redundant operators can be reabsorbed by a change of variables and are therefore not physically relevant. This however cannot be applied as such when testing whether a Ward identity is valid or not. Indeed, this would lead us to the absurd conclusion that a generic theory with two scalar fields can always be made O(2)-invariant by reabsorbing the redundant operator appearing in the right-hand-side of Eq.~(\ref{eq_o2}) through a field redefinition.}
		
		The situation is however different in the case of conformal invariance. Assume indeed that we find a redundant, integrated vector operator of dimension -1. Such an operator would  be of the form \footnote{We focus on the Ising case but the same discussion generalizes to other universality classes.} $\int d^dx (\delta \Gamma_k)/(\delta \phi (x))\mathcal O_\mu(x)$ where the operator $\mathcal O_\mu$ depends on $x$ only through the field argument (as explained in Sect.~\ref{Sec.II}, an explicit $x$-dependence would be inconsistent with the translational invariance of the operator $\Sigma_\mu$). This operator would yield a potential violation of the Ward identity:
		\begin{equation}
		\label{redundant_break}
		\int_{x} (K_{\mu}^{x}-2D_\star^{\phi}x_{\mu})\phi(x)\frac{\delta\Gamma_{k}}{\delta\phi(x)}-\frac{1}{2}\int_{x,y}\partial_{t}R_{k}(|x-y|)(x_{\mu}+y_{\mu})G_{k}(x;y)=\int_x \frac{\delta \Gamma_k}{\delta \phi (x)}\mathcal O_\mu(x)
		\end{equation}
		The right-hand-side could actually be reabsorbed in a modification of the conformal transformation of the field $\phi$. \cite{Nakayamathanks} At odds with the case of internal symmetries, the modified conformal transformation $\phi(x)\to \phi(x)+\epsilon_\mu [(K_{\mu}^{x}-2D_\star^{\phi}x_{\mu})\phi(x)-\mathcal O_\mu(x)]$ is nontrivial because the $x$-independent term $\mathcal O_\mu$ cannot compensate the usual variations and the bracket is therefore non zero.  
		
		It can be shown that the modified conformal transformation, together with the usual translation, rotation, and scale transformation satisfy the conformal algebra.
		
		To conclude, the existence of a putative redundant, integrated vector operator of dimension -1 would not lead to a breaking of conformal invariance but, instead, to a modification of the transformation of the field.
		
		\subsection{Exact scaling dimensions of some redundant operators}
		
		In this section, we show that some redundant operators have simple scaling dimensions. We work in the framework of NPRG described in Sect.~\ref{secNPRG}.
			We can choose the Hamiltonian (or action) to be of the
			Ginzburg-Landau type:
			\begin{equation}
			\label{eq_action_micro}
			S[\phi]=\int_x \frac{1}{2}(\nabla \phi)^2+\frac{1}{2}r_\Lambda\phi^2+\frac{u_\Lambda}
			{4!}\phi^4,
			\end{equation}
			where $\int_x=\int d^d x$. In order to determine the scaling dimension of an operator, we study the evolution of the corresponding coupling under the renormalization-group flow in the vicinity of the fixed point. To this end, we add to the action a part which couples to a vector operator:
			\begin{equation}
			S_{\text V}[\phi]=\int_x \frac{a^\mu_\Lambda}{3!}\phi^3\partial_\mu\Delta\phi.
			\end{equation}
			Up to integrations by parts, this operator is the same as the one considered in \cite{Meneses:2018xpu}. Moreover, it has been proved to be the most relevant integrated vector operator invariant under $\mathbb{Z}_2$ symmetry near $d=4$ \cite{delamotte2016scale}.

		We perform an infinitesimal transformation of the integration variable: $\phi\to\phi-a^\mu_\Lambda/u_\Lambda \partial_\mu\Delta \phi$ in the path integral appearing in Eq.~(\ref{regulatedgeneratingfunc}). It is readily found that the quadratic pieces in the action, including the regulating term $\Delta S_k$, are invariant under this transformation. The variation of the quartic part of the action is found to compensate exactly $S_{\text V}$. We thus find that
			\begin{equation}
			W_{k}[J,a^\mu_\Lambda]=W_{k}[J+\frac{a^\mu_\Lambda}{u_\Lambda}\partial_\mu\Delta J,0]+\mathcal O(a^\mu_\Lambda a^\nu_\Lambda)
			\end{equation}
			At the level of the effective average action, this relation implies
			\begin{equation}
			\Gamma_{k}[\phi,a^\mu_\Lambda]=\Gamma_k[\phi+a^\mu_\Lambda\partial_\mu \Delta\phi,0]+\mathcal O(a^\mu_\Lambda a^\nu_\Lambda
			).
			\end{equation}
			This last equation states that the evolution of the effective action with an infinitesimal $a^\mu_\Lambda$ is related to the effective action at vanishing $a^\mu_\Lambda$, up to a modification of the field. This can be used in the following way. Defining the running coupling constants $u_k$ and $a^\mu_k$ as the prefactors of, respectively, $\int_x \frac{1}{4!}\phi^4$ and $\int_x \frac{1}{3!}\phi^3\partial_\mu\Delta\phi$ in $\Gamma_k$, we obtain that $a^\mu_k/u_k$ is constant along the flow. To obtain the scaling dimension of the vector operator, we introduce dimensionless, renormalized quantities (denoted with tilde) as
			\begin{align}
			\tilde x&=k x\\
			\tilde \phi(\tilde x)  &=k^{-(d-2)/2}Z_k^{1/2}\phi(x),
			\end{align}
			where $Z_k$ scales as $ k^{-\eta}$ at the Wilson-Fisher fixed point with $\eta$ the anomalous dimension. 
			The renormalized coupling constants are thus: 
			\begin{align}
			\tilde u_k&=k^{d-4} Z_k^{-2} u_k\\
			\tilde a_k^\mu&=k^{d-1} Z_k^{-2}  a_k^\mu.
			\end{align}
			At the critical point, $\tilde u$ flows to a fixed point value $u_\star$. Consequently, when $k\to 0$,
			\begin{equation}
			\tilde a_k^\mu\sim a_\Lambda^\mu\frac{u_\star}{u_\Lambda}k^3
			\end{equation}
			which shows that the scaling dimension of $a^\mu$ is exactly 3.
		
		The proof given above relies strongly on the particular microscopic action given in Eq.~(\ref{eq_action_micro}). This gives interesting non-universal information on the flow of the coupling $a_k^\mu$, but confers a preeminent role to the peculiar form of the Hamiltonian. To overcome this issue, we now present an alternative proof of the same result, but which is based on the exact flow equation \ref{wettericheq} for the effective average action, expressed in terms of dimensionless, renormalized, fields:
			\begin{equation}
			\begin{split}
			\partial_t \Gamma_k[\tilde \phi]=&\int_{\tilde x} \frac{\delta \Gamma_k}{\delta \tilde \phi(\tilde x)}\left(\tilde x^\rho\partial_{\tilde x^\rho}-\alpha\right)\tilde \phi(\tilde x)+\frac 12\int_{\tilde x\tilde y}\partial_t\tilde R(\tilde x-\tilde y)\tilde P_k(\tilde x,\tilde y)
			\label{eq_flow2}
			\end{split}
			\end{equation}
			\vspace{.5cm}
			where $ R_k( x)=Z_k k^{d+2}\tilde R(k x)$, $\alpha=-(d-2+\eta)/2$, $t=\log(k/\Lambda)$ and $\tilde P_k$ is the dimensionless, renormalized, propagator:
			\begin{equation}
			\int_{\tilde y}    \tilde P_k(\tilde x,\tilde y)\left[\frac{\delta^2\Gamma_k}{\delta\tilde \phi(\tilde y)\delta\tilde \phi(\tilde z)}+\tilde R(\tilde y-\tilde z)\right]=\delta(\tilde x-\tilde z)
			\end{equation}
		
		We now identify an exact eigenvector of the linearized flow. To this end, we add to the Wilson-Fisher fixed-point effective action $\Gamma_\star$ a small perturbation
			\begin{equation}
			\label{eq_pert}
			\Gamma_k=\Gamma_\star+\tilde r_\mu(t)\int_{ \tilde x}
			\frac{\delta\Gamma_\star}{\delta \tilde \phi(\tilde x)} \tilde
			\partial_\mu\tilde\Delta \tilde \phi(\tilde x)
			\end{equation}
			and we compute the flow of this functional at linear order in $\tilde r_\mu$.
			\begin{equation}
			\begin{split}
			\partial_t{\tilde r}_\mu(t) \int_{\tilde x}\frac{\delta\Gamma_\star}{\delta \tilde \phi(\tilde x)} \tilde{\partial}_\mu&\tilde{\Delta} \tilde \phi(\tilde x)=
			\tilde r_\mu(t)\int_{\tilde x\tilde y} \frac{\delta \ }{\delta \tilde \phi(\tilde y)}\left[\frac{\delta\Gamma_\star}{\delta \tilde \phi(\tilde x)} \tilde\partial_\mu\tilde\Delta \tilde \phi(\tilde x)\right]\left(\tilde y^\rho\partial_{\tilde y^\rho}-\alpha\right) \tilde \phi(\tilde y)\\&-\frac12 \tilde r_\mu(t)\int_{\tilde x\tilde y\tilde z\tilde t\tilde w}\partial_t\tilde R(\tilde x-\tilde y)\tilde P_\star(\tilde y,\tilde z)\Gamma_\star^{(3)}(\tilde z,\tilde t,\tilde w)\tilde P_\star(\tilde w,\tilde x)\tilde\partial_\mu\tilde\Delta\tilde \phi(\tilde t)
			\end{split}
			\end{equation}
			On the other hand, if we derive the fixed point equation with respect to $\tilde \phi(\tilde x)$, multiply by $ \tilde\partial_\mu\tilde\Delta \tilde \phi(\tilde x)$ and integrate over $\tilde x$, we get
			\begin{equation}
			\begin{split}
			0=\int_{\tilde x\tilde y}\tilde\partial_\mu\tilde\Delta \tilde \phi(\tilde x)&\frac{\delta \ }{\delta \tilde \phi(\tilde x)}\left[\frac{\delta\Gamma_\star}{\delta \tilde \phi(\tilde y)} \left(\tilde y^\rho\partial_{\tilde y^\rho}-\alpha\right)\tilde \phi(\tilde y)\right]\\&-\frac12 \int_{\tilde x\tilde y\tilde z\tilde t\tilde w}\partial_t\tilde R(\tilde x-\tilde y)\tilde P_\star(\tilde y,\tilde z)\Gamma_\star^{(3)}(\tilde z,\tilde t,\tilde w)\tilde P_\star(\tilde w,\tilde x)\tilde\partial_\mu\tilde\Delta\tilde \phi(\tilde t)
			\end{split}
			\end{equation}
			Combining the two equations, we obtain:
			\begin{equation}
			\begin{split}
			\partial_t {\tilde r}_\mu(t) \int_{\tilde x} \frac{\delta\Gamma_\star}{\delta \tilde\phi(\tilde x)} \tilde\partial_\mu\tilde\Delta^n \tilde\phi(\tilde x)=
			\tilde r_\mu(t)\int_{\tilde x\tilde y} \frac{\delta\Gamma_\star}{\delta \tilde \phi(\tilde x)} \left[\tilde \partial_\mu\tilde \Delta ,\tilde x^\rho\partial_{\tilde x^\rho}\right] \tilde\phi(\tilde x)
			\end{split}
			\label{proof_EV}
			\end{equation}
			The commutator is easily evaluated to be equal to $3\tilde \partial_\mu\tilde \Delta $. From this we deduce that the small perturbation introduced in Eq.~(\ref{eq_pert}) is an exact eigenoperator of the flow around the fixed point, with eigenvalue 3. This is consistent with the result found in the one-loop calculation of \cite{delamotte2016scale}.
		
		We can generalize the previous result in different ways. First, we can change the power of the Laplacian in Eq.~(\ref{eq_pert}) from unity to a positive integer $n$. The main change appears at the level of Eq.~(\ref{proof_EV}), where the commutator is now $[\tilde\partial_\mu\tilde\Delta^n ,\tilde x^\rho\partial_{\tilde x^\rho}]=(2n+1)\tilde\partial_\mu\tilde\Delta^n$. This implies that the associated eigenvector has dimension $2n+1$. As a check of this result, we have considered the vector eigenoperators compatible with the $\mathbb{Z}_2$ symmetry whose scaling dimensions are 5 in $d=4$ and we have computed their first correction in $\epsilon=4-d$. There are four (independent) such operators: one ($O_6^3$) with 6 powers of the field and 3 derivatives and three ($O_{4,i}^5$ with $i\in\{1,2,3\}$) with 4 powers of the field and 5 derivatives. A one-loop calculation shows that $O_6^3$ has scaling dimension $5-5\epsilon/3+\mathcal O(\epsilon^2)$. The eigenvectors $O_{4,i}^5$ have dimensions $5+\mathcal O(\epsilon^2)$, $5-4\epsilon/9+\mathcal O(\epsilon^2)$ and $5-2\epsilon/3+\mathcal O(\epsilon^2)$. The eigenoperator with scaling dimension $5+\mathcal O(\epsilon^2)$ is found to be $\int_{\tilde x} \tilde \phi^3 \tilde \partial_\mu\tilde \Delta^2\tilde \phi$, in agreement with the general result mentioned above. Other relations can be obtained if we consider in Eq. (\ref{eq_pert}) an odd number of derivatives, with Lorentz indices not necessarily contracted together.
		
		The present result also generalizes to the long-range Ising model, where the interaction between spins is not limited to nearest neighbors but decay as a power-law:
			\begin{equation}
			H=-\sum_{i,j}J(i-j)S_i S_j
			\end{equation}
			where $J(i-j)\sim |i-j|^{-d-\sigma}$ and $\sigma$ is the exponent characterizing the decrease of the interactions. When $0<\sigma<2-\eta$, the model still has an
			extensive free-energy but belongs to a different universality class than the local Ising model. The Ginzburg-Landau Hamiltonian is identical to the one given in Eq.~(\ref{eq_action_micro}) except that the quadratic part is now, in Fourier space,
			\begin{equation}
			\int \frac{d^dq}{(2\pi)^d}\phi(-q) q^{\sigma}\phi(q).
			\end{equation}
			It is easy to verify that all the present analysis still applies to this case. We have checked that the one-loop calculation around the upper critical dimension
			$d_c=2\sigma$ gives that the most relevant integrated vector operator has scaling dimension $3+\mathcal{O}(\epsilon^2)$. This result is important because
			it justifies the use of the conformal bootstrap program in this model \cite{Paulos:2015jfa}, at least near the upper critical dimension.
		
		We can also generalize the result to other internal groups. For $O(N)$ theories, an exact eigenoperator can be found by adding a common O($N$) index on both the functional derivative and the field appearing in Eq.~(\ref{eq_pert}) and summing over this index. The associated eigenvalue is again 3 (or $2n+1$, if we change the power of the Laplacian). In \cite{delamotte2016scale} [see also Eq.~(\ref{eq:fpfeEps})], we computed the scaling dimensions of the two vector operators of lowest dimension in an expansion in $\epsilon$ and found $3+\mathcal{O}(\epsilon^2)$ and $3-6\epsilon/(N+8)+\mathcal{O}(\epsilon^2) $. This result is consistent with the nonrenormalization theorem proven here. Let us stress, however, that in the $O(N)$ model the non renormalization theorem does not constraint the leading vector operator but the next-to-leading, as can be seen already at one-loop level \cite{delamotte2016scale}.
		
		For completeness, we give other exact eigenvectors which can be obtained following the same idea. These are, however, not invariant under the internal symmetries of the theory, involve even derivatives and are not relevant to our discussion on conformal invariance. A well-known example is the eigenvector associated with the external magnetic field: $\int_{ \tilde x}
			\frac{\delta\Gamma_\star}{\delta \tilde \phi(\tilde x)}$ with eigenvalue $\alpha=-(d-2+\eta)/2$. 
			Another exact eigenvector, valid for the O$(N)$ model is  $\int_{ \tilde x}
			\epsilon^{ab}\frac{\delta\Gamma_\star}{\delta \tilde \phi^b(\tilde x)} \tilde\Delta^n \tilde \phi^a(\tilde x)$
			where $\epsilon^{ab}$ is antisymmetric, which is associated with the eigenvalue $2n$.
		
		It was shown in ref.~\cite{Meneses:2018xpu} that there exists other redundant operators in $d=4-\epsilon$. In fact, all local operators of dimension up to 9 are redundant modulo total derivatives. Equivalently, all integrated vector operators of dimension up to 5 are redundant. For several of these, we were however not able to determine analytically their scaling dimensions.
		
		\section{$\epsilon$-expansion}\label{Ap:EpsExp}
		
		In this appendix, we describe the 1-loop calculation of the two lowest scaling
		dimensions of vector operators. We present here the calculation
		in the framework of the NPRG, but of course, the calculation could be
		performed within more standard approaches, such as the Minimal
		Substraction scheme. 
		
		We start by observing that at tree-level (zero loop) the
		effective action takes its bare form:
		\begin{align}
		\Gamma^{(tree)}_{k}=S=\int_x\Big\{r \rho+\frac{1}{3!}u\rho^2+ \frac{1}{4}a_{\mu}\partial_{\mu}\rho\partial_{\nu}\phi_{i}\partial_{\nu}\phi_{i}
		+\frac{1}{2}b_{\mu}\partial_{\mu}\phi_{i}\partial_{\nu}\phi_{i}\partial_{\nu}\rho\Big\}
		\end{align}
		Differentiating successively with respect to
		$\phi_{n_{i}}\left(x_{i} \right)$ and Fourier transforming, we obtain
		the form of the non-zero vertices:
		\begin{align}
		&{\Gamma_{i_{1}i_{2}}^{(2,tree)}}(p_1) = \delta_{i_1i_2}(r+p_1^2)\\
		&{\Gamma_{i_{1}i_{2}i_{3}i_{4}}^{(4,tree)}}(p_1,p_2,p_3)
		= \frac{u}{3}\Big[\delta_{i_{1}i_{2}}\delta_{i_{3}i_{4}}+\delta_{i_{1}i_{3}}\delta_{i_{2}i_{4}}+\delta_{i_{1}i_{4}}\delta_{i_{2}i_{3}}\Big]\nonumber\\ 
		&+i \frac{a_{\mu}-b_\mu}{2}\Big[\left(p_{1}+p_{2}\right)^{\mu}\big(p_{1}\cdot p_{2}- p_{3}\cdot p_{4}\big)\delta_{i_{1}i_{2}}\delta_{i_{3}i_{4}}
		+ \left(p_{1}+p_{3}\right)^{\mu}\big(p_{1}\cdot p_{3}- p_{2}\cdot p_{4}\big)\delta_{i_{1}i_{3}}\delta_{i_{2}i_{4}}\nonumber \\ 
		&+  \left(p_{1}+p_{4}\right)^{\mu}\big(p_{1}\cdot p_{4}- p_{2}\cdot p_{3}\big)\delta_{i_{1}i_{4}}\delta_{i_{2}i_{3}}\Big]\nonumber   \\
		&-i\frac{b_{\mu}}{2}
		\left(p_{1}^{\mu}p_{1}^2+p_{2}^{\mu}p_{2}^2+p_{3}^{\mu}p_{3}^2+p_{4}^{\mu}p_{4}^2\right)\left(\delta_{i_{1}i_{2}}\delta_{i_{3}i_{4}}
		+\delta_{i_{1}i_{3}}\delta_{i_{2}i_{4}}+\delta_{i_{1}n_{i}}\delta_{i_{2}i_{3}}\right) \label{tree-level}
		\end{align}
		In the previous equations, we have omitted for notation simplicity
		the index $k$ on the coupling constants. The momentum $p_4$ in $\Gamma^{(4)}$ is fixed by
		momentum conservation: $\sum_{i=1}^{n} p_i=0$. The flow of $a_{\mu}$
		and $b_{\mu}$ are deduced from the flow equation of $\Gamma^{(4)}$ at
		zero external field, which is obtained by differentiating four times the 
		RG equation (\ref{wettericheq}) and evaluating it at
		$\phi=0$. We obtain:
		\begin{align}
		\label{eq_flowG4}
		\partial_{t}&\Gamma^{(4)}_{i_{1}i_{2}i_{3}i_{4}}(p_1,p_2,p_3)
		=\sum_{il}\int_{q}\partial_t R_k\left(q^2\right)G_k^2\left(q^2\right)\bigg(-\frac 12 \Gamma^{(6)}_{k,k,i_{1},i_{2},i_{3},i_{4}}(q,-q,p_1,p_2,p_3)\nonumber \\
		&+ G_k\left((q+p_{1}+p_{2})^2\right)
		\Gamma^{(4)}_{ki_{1}i_{2}l}(q,p_1,p_2)\Gamma^{(4)}_{li_{3}i_{4}k}(q+p_1+p_2,p_3,p_4)+\text{2 perms.}\bigg).
		\end{align}
		At one loop, we can neglect the first term and replace the vertices in the
		right hand side of the flow equation by its tree-level form (\ref{tree-level}).
		Summing over $i$ and $l$ in the product of the $\Gamma_k^{(4,tree)}$, and keeping only terms which are at most linear in $a_\mu$ and $b_\mu$,  we
		find that the result is independent of $q$:
		\begin{equation}
		\begin{split}
		&{\Gamma_{ki_{1}i_{2}l}^{(4,tree)}}(q,p_1,p_2){\Gamma_{li_{3}i_{4}k}^{(4,tree)}}(-q,p_3,p_4)+2 \text{perms.}
		\\ &= \frac{u^{2}}{9}\left(N+8\right)\big(\delta_{i_{1}i_{2}}\delta_{i_{3}i_{4}}+\delta_{i_{1}i_{3}}\delta_{i_{2}i_{4}}+\delta_{i_{1}i_{4}}\delta_{i_{2}i_{3}}\big)\\
		&+\frac{iua_{\mu}}{6}\Big\{\delta_{i_{1}i_{2}}\delta_{i_{3}i_{4}}\big[ (p_{1}+p_{2})^{\mu}(p_{1}\cdot p_{2}-p_{3}\cdot p_{4})\left(N+2\right)
		-2\sum_{k=1}^{4}(p_{k}^{\mu}p_{k}^{2})\big]+2 \text{ perms.}\Big\}\\
		&-\frac{iub_{\mu}}{6}\Big\{\delta_{i_{1}i_{2}}\delta_{i_{3}i_{4}}
		\big[ \left(p_{1}+p_{2}\right)^{\mu}(p_{1}\cdot p_{2}-p_{3}\cdot p_{4})\left(N+2\right)
		+\sum_{k=1}^{4}(p_{k}^{\mu}p_{k}^{2})\left(N+6\right)\big] +2 \text{ perms.}\Big\}.
		\end{split}
		\label{eq_sommation_G4}
		\end{equation}
		where cyclic permutations of the indices 2, 3 and 4 are not written explicitly.
		
		The next step consists in identifying the prefactors of a given
		structure which involve both vector indices and momenta [see Eq.~(\ref{tree-level})] in the left hand side  and right hand side of the flow equation
		(\ref{eq_flowG4}). This implies that we must expand the r.h.s in
		powers of the external momenta and extract terms of order zero and order three in momenta.
		However, Eq.~(\ref{eq_sommation_G4}) shows that the product of vertices already has a contribution with 0 and 3 powers of the external momenta.
		As a consequence, we can put, in
		the propagator, the external momenta to zero. 
		
		We can now extract the flows of $a_\mu$ and
		$b_\mu$, which read:
		\begin{align}
		\partial_{t}u&=\frac{\left(N+8\right)}{3}u^{2}\int_{q}{\dot{R}\left(q\right)G^{3}\left(q\right)}\nonumber\\
		\partial_{t}a_{\mu}&=\frac{u}{3}\left[ \left(N+4\right)a_{\mu}+4b_{\mu} \right]\int_{q}{\dot{R}\left(q\right)G^{3}\left(q\right)}\nonumber\\
		\partial_{t}b_{\mu}&=\frac{u}{3}\left[ 2a_{\mu}+\left(N+6\right)b_{\mu} \right]\int_{q}{\dot{R}\left(q\right)G^{3}\left(q\right)}.
		\end{align}
		The flow equations for the dimensionless variables are easily derived and correspond to those given in Eq.~(\ref{eq:feEps}).

		\section{Large N Expansion}\label{Ap:LargeN}
		We discuss in this appendix the large-$N$ calculation of the two smallest scaling
		dimensions of the vector operators (which tend to 3 when $d\to4$) as well as another one, (which tend to 5 when $d\to4$).  These can be deduced from the
		calculation of $\Gamma^{\left(2\right)}$, $\Gamma^{\left(4\right)}$ and $\Gamma^{\left(6\right)}$
		at vanishing external field in the large $N$ limit. We recall that the
		large $N$ limit is performed at fixed $\hat u=u N$,
		$\hat{a}_\mu=a_\mu N$, $\hat{b}_\mu=b_\mu N$ and $\hat{c}_\mu=c_\mu N^2$. Moreover, we only
		need the flow of $\hat{a}_\mu$, $\hat{b}_\mu$ and $\hat{c}_\mu$ at linear order in
		$\hat{a}_\mu$, $\hat{b}_\mu$ and $\hat{c}_\mu$.
		\begin{equation}
		{\int_x \left\{\frac{\hat u^\Lambda}{4!N}\phi_{i}^{2}\phi_{j}^{2}
			+\frac{\hat{a}^\Lambda_{\mu}}{4N}\phi_{i}\partial_{\mu}\phi_{i}\partial_{\nu}\phi_{j}\partial_{\nu}\phi_{j}
			+\frac{\hat{b}^\Lambda_{\mu}}{2N}\phi_{i}\partial_{\nu}\phi_{i}\partial_{\nu}\phi_{j}\partial_{\mu}\phi_{j}
			+\frac{\hat{c}^\Lambda_{\mu}}{4N^2}\phi_{i}\partial_{\nu}\phi_{i}\phi_{j}\partial_{\nu}\phi_{j}\phi_{l}\partial_{\mu}\phi_{l}\right\}}
		\end{equation}
		
		The bare propagators and 4-point vertex are easily deduced from those
		given in Eq.~(\ref{tree-level}). The 6-point vertex is given by:
		\begin{align}
		&{\Gamma_{i_{1}i_{2}i_{3}i_{4}i_{5}i_{6}}^{(6,tree)}}(p_1,p_2,p_3,p_4,p_5)
		= -i\frac{c_{\mu}^{\Lambda}}{2}\bigg\lbrace\delta_{i_{1}i_{2}}\delta_{i_{3}i_{4}}\delta_{i_{5}i_{6}}\label{c-tree-level}\\ \nonumber
		&\Big[\left(p_{1}+p_{2}\right)^{\mu}(p_{1}+ p_{2})^2+\left(p_{3}+p_{4}\right)^{\mu}(p_{3}+ p_{4})^2+\left(p_{5}+p_{6}\right)^{\mu}(p_{5}+ p_{6})^2\Big]+perm\bigg\rbrace\nonumber               
		\end{align}
		where {\it perm} represents the 14 permutations which lead to different  combinations of Kronecker delta.
		
		\subsection{A source of simplification}
		\label{sect_simpl}

		A major simplification occurs in the calculation, which is a
		consequence of the following property: the contribution linear in
		$a_\mu$ or $b_\mu$ in the 4-point vertex  which is proportional to $\delta_{i_{1}i_{2}}\delta_{i_{3}i_{4}}$ vanishes in the exceptional
		configurations where the momenta are opposite by pairs in different delta's (that
		is, if $p_1+p_3=p_2+p_4=0$ or if $p_1+p_4=p_2+p_3=0$). As a consequence, in a diagram made of a chain of bubbles
		(see Fig.~\ref{fig_3}), if a $\Gamma^{(4)}$ connecting two bubbles is replaced by a
		perturbation $a_\mu$ or $b_\mu$, the diagram vanishes. Otherwise stated, in a chain of bubbles diagram,
		the perturbations $a_\mu$ and $b_\mu$ only occur when attached to an
		external leg (see Fig.~\ref{fig_4.5}).
		
		A somewhat similar situation occurs when a $c_\mu$ perturbation appears in a chain of bubble for the 4-point function. When this perturbation is inserted in a chain (whether in contact or not with external legs), given the structure of Eq.~\eqref{c-tree-level}, the legs from the isolated loop cancels by itself and the remaining two pairs cancel with each other. This closely resembles the property mentioned previously that $a_\mu$ and $b_\mu$ don't appear in an inner vertex of the chains. Moreover, the specific momentum structure appearing in the $c_\mu$ vertex, implies that it can't appear neither attached to an external leg. We conclude that $c_\mu$ does not contribute to the flow of $a_\mu$ and $b_\mu$ at leading order in $1/N$.
		
		The situation is even simpler in the calculation of $\Gamma^{(2)}$
		because, by conservation of the momenta, the external legs have opposite momenta.
		In this situation, the perturbation 4-point vertex does
		not even contribute when attached to the external legs. As a
		consequence, the cactus diagrams for $\Gamma^{(2)}$ are independent of
		$a_\mu$ and $b_\mu$. This result is important because it implies that
		the inverse full propagator (which re-sums all cactus diagrams for
		$\Gamma^{(2)}$) is independent of $a_\mu$, $b_\mu$ and $c_\mu$.

		\subsection{Computation of $\Gamma^{(2)}$}
		We first discuss the (standard) calculation of $\Gamma^{(2)}$ at
		leading order. As discussed above, we can remove $a_\mu$ and $b_\mu$
		from this calculation. The sum of the cactus diagrams shown in
		Fig.~\ref{fig_1} leads to
		\begin{equation}
		\Gamma_{ij}^{(2)}(p)=\delta_{ij}\Big\{p^2+r^\Lambda
		+\frac{\hat{u}^\Lambda}{6}\int_q\frac{1}{q^2+r^\Lambda+\overline{\Sigma}\left(r^\Lambda\right)}\Big\}=\delta_{ij}\Big\{p^2+r\Big\}
		\end{equation}
		\noindent where $r=r^{\Lambda}+\overline{\Sigma}\left(r^\Lambda\right)$ and
		$\overline{\Sigma}\left(r^\Lambda\right)$ satisfies the gap equation:
		\begin{equation}
		\overline{\Sigma}\left(r^\Lambda\right)=\frac{\hat{u}^\Lambda}{6}\int_q\frac{1}{q^2+r^\Lambda+\overline{\Sigma}\left(r^\Lambda\right)}.
		\end{equation}
		As is well known, the only effect of the cactus diagrams is to
		modify the mass.
		
		\subsection{Computation of $\Gamma^{\left(4\right)}$}
		
		In contrast to ${\Gamma}^{\left(2\right)}$, the
		${\Gamma}^{\left(4\right)}$ vertex has corrections linear in
		$a^\Lambda_{\mu}$ and $b^\Lambda_{\mu}$ to leading order in the
		$\frac{1}{N}$ expansion (i.e. to order $\frac{1}{N}$). It is
		convenient to decompose the 4-point vertex function as
		$\Gamma^{\left(4\right)}=\Gamma^{\left(4\right)}_u+\Gamma^{\left(4\right)}_{a_{\mu}}+\Gamma^{\left(4\right)}_{b_{\mu}}$,
		where the first term is independent of $a_\mu$ and $b_\mu$, the second
		term is linear in $a_\mu$ and the third is linear in $b_\mu$. We omit
		all other terms which do not enter into the calculation of the scaling
		dimensions we are interested in.
		
		The term $\Gamma^{\left(4\right)}_u$ is the simplest one since it
		corresponds to the usual theory with
		$a^\Lambda_{\mu}=b^\Lambda_{\mu}=0$. This gives the standard large-$N$
		result:
		\begin{align}
		\Gamma^{\left(4\right)}_{u,i_1,i_2,i_3,i_4}(p_1,p_2,p_3)=\frac{\hat{u}^\Lambda}{3N}\left[\frac{\delta_{i_{1}i_{2}}\delta_{i_{3}i_{4}}}{1+\frac{\hat{u}^\Lambda}{6}\Pi\left(p_{1}+p_{2}\right)}\right.
		&+\frac{\delta_{i_{1}i_{3}}\delta_{i_{2}i_{4}}}{1+\frac{\hat{u}^\Lambda}{6}\Pi\left(p_{1}+p_{3}\right)}\nonumber\\
		&+\left.\frac{\delta_{i_{1}i_{4}}\delta_{i_{2}i_{3}}}{1+\frac{\hat{u}^\Lambda}{6}\Pi\left(p_{1}+p_{4}\right)}\right]
		\end{align}
		where the function $\Pi\left(p\right)$ is defined as:
		\begin{equation}
		\Pi\left(p\right)=\int_q \frac{1}{q^2+r}\frac{1}{\left( q+p\right)^2+r}
		\end{equation}
		
\begin{figure}[!ht]
			\begin{center}
				\includegraphics[width=1\textwidth]{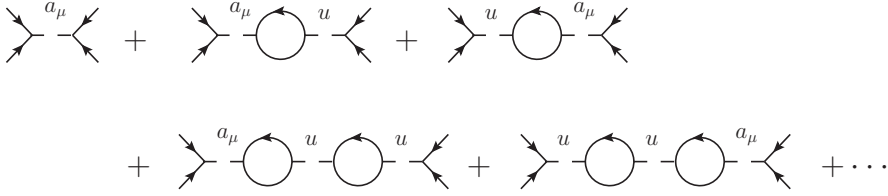}
			\end{center}
			\caption{Diagrams contributing to $\Gamma^{\left(4\right)}_{a_\mu}$.}
			\label{fig_4.5}
		\end{figure}

		We now consider $\Gamma^{\left(4\right)}_{a_{\mu}}$. For simplicity,
		we focus on the contribution proportional to
		$\delta_{i_1i_2}\delta_{i_3i_4}$. The other contribution are obtained
		by permutations of the external legs. The set of diagrams which
		contribute is easy to characterize because the perturbation ($a_\mu$
		in this case) must be attached to the external legs (see
		Sect.~\ref{sect_simpl}). The chain of bubbles diagrams for
		$\Gamma^{\left(4\right)}_{a_{\mu}}$ are depicted in
		Fig.~\ref{fig_4.5}.
		
		The diagram with $n$ couplings $\hat{u}^\Lambda$ and one $\hat{a}^\Lambda_{\mu}$ connected to $p_{1}$ and $p_{2}$ is equal to:
		\begin{equation}
		\frac{i\hat{a}^\Lambda_{\mu}}{2N}\int_q\frac{\left(p_{1}+p_{2}\right)^{\mu}\left[p_{1}\cdot p_{2}+q\cdot\left(q+p_{1}+p_{2}\right)\right]}
		{\left[q^{2}+r\right]\left[\left(q+p_{1}+p_{2}\right)^{2}+r\right]}\left(\frac{-\hat{u}^\Lambda}{4!}\right)^{n}\frac{4^{n}n!C^{n+1}_{n}}{\left(n+1\right)!}
		\left(\Pi\left(p_{1}+p_{2}\right)\right)^{\left(n-1\right)},
		\end{equation}
		If $\hat{a}^\Lambda_{\mu}$ is connected to $p_{3}$ and $p_{4}$, we get:
		\begin{equation}
		\frac{i\hat{a}^\Lambda_{\mu}}{2N}\int_q\frac{\left(p_{1}+p_{2}\right)^{\mu}\left[-q\cdot\left(q+p_{1}+p_{2}\right)-p_{3}\cdot p_{4}\right]}
		{\left[q^{2}+r\right]\left[\left(q+p_{1}+p_{2}\right)^{2}+r\right]}\left(\frac{-\hat{u}^\Lambda}{4!}\right)^{n}\frac{4^{n}n!C^{n+1}_{n}}{\left(n+1\right)!}\left(\Pi\left(p_{1}+p_{2}\right)\right)^{\left(n-1\right)}.
		\end{equation}
		When adding both diagrams we get the result for $n$ couplings $\hat{u}^\Lambda$ and one $\hat{a}^\Lambda_{\mu}$:
		\begin{equation}
		\label{eq_intermamu}
		\frac{i\hat{a}^\Lambda_{\mu}}{2N}\left(-\frac{\hat{u}^\Lambda}{6}\Pi\left(p_{1}+p_{2}\right)\right)^{n}\left(p_{1}+p_{2}\right)^{\mu}\left[p_{1}\cdot p_{2}-p_{3}\cdot p_{4}\right].
		\end{equation}
		Note that the previous construction does not make sense for
		$n=0$. However, it happens that Eq.~(\ref{eq_intermamu}) evaluated at
		$n=0$ indeed represents the contribution of the first diagram of
		Fig.~\ref{fig_4.5} with one $a_\mu$ and no $\hat u$. 
		It is straightforward to sum this general expression for all $n$ to
		get:
		\begin{equation}
		\Gamma^{\left(4\right)}_{a_{\mu}}=\frac{i\hat{a}^\Lambda_{\mu}}{2N\left(1+\frac{\hat{u}^\Lambda}{6}\Pi\left(p_{1}+p_{2}\right)\right)}
		\left(p_{1}+p_{2}\right)^{\mu}\left[p_{1}\cdot p_{2}-p_{3}\cdot p_{4}\right]\delta_{i_1i_2}\delta_{i_3i_4}+\text{2 perms.}
		\end{equation}

		The calculation for $\Gamma_{b_\mu}^{(4)}$ proceeds in the same way. The
		contribution of diagrams with one $b_\mu$ and $n$ couplings $\hat u$ (again
		focusing on the contribution proportional to
		$\delta_{i_1i_2}\delta_{i_3i_4}$) is:
		\begin{equation}
		\frac{-i\hat{b}^\Lambda_{\mu}}{2N}\left(-\frac{\hat{u}^\Lambda}{6}\Pi\left(p_{1}+p_{2}\right)\right)^{n}\left[p_{1}^{\mu}p_{1}^{2}+p_{2}^{\mu}p_{2}^{2}+p_{3}^{\mu}p_{3}^{2}+p_{4}^{\mu}p_{4}^{2}+\left(p_{1}+p_{2}\right)^{\mu}(p_{1}\cdot p_{2}-p_{3}\cdot p_{4})\right].
		\end{equation}
		To sum up,  the four-point vertex with at most one $a_\mu$
		or one $b_\mu$ is
		\begin{equation}
		\begin{split}
		\Gamma^{\left(4\right)}_{i_1i_2i_3i_4}(p_1,p_2,p_3)=&\frac{\delta_{i_{1}i_{2}}\delta_{i_{3}i_{4}}}{N\big(1+\frac{\hat{u}^\Lambda}{6}\Pi\left(p_1+p_2\right)\big)}
		\bigg\lbrace\frac{\hat{u}^\Lambda}{3}+i \frac{\hat{a}^\Lambda_{\mu}-\hat{b}^\Lambda_{\mu}}{2}\left(p_{1}+p_{2}\right)^{\mu}\left[p_{1} \cdot p_{2} - p_{3} \cdot p_{4}\right] \\
		& -i
		\frac{\hat{b}^\Lambda_{\mu}}{2}\left(p_{1}^{\mu}p_{1}^2+p_{2}^{\mu}p_{2}^2+p_{3}^{\mu}p_{3}^2+p_{4}^{\mu}p_{4}^2\right)\bigg\rbrace+\text{2 perms.}    
		\end{split}
		\end{equation}
		where, again, the permutations are obtained by a cyclic permutation of
		the external indices 2, 3 and 4.
		
		\subsection{Computation of $\Gamma^{\left(6\right)}$}
		
\begin{figure}[!ht]
			\begin{center}
				\includegraphics[width=.5\textwidth]{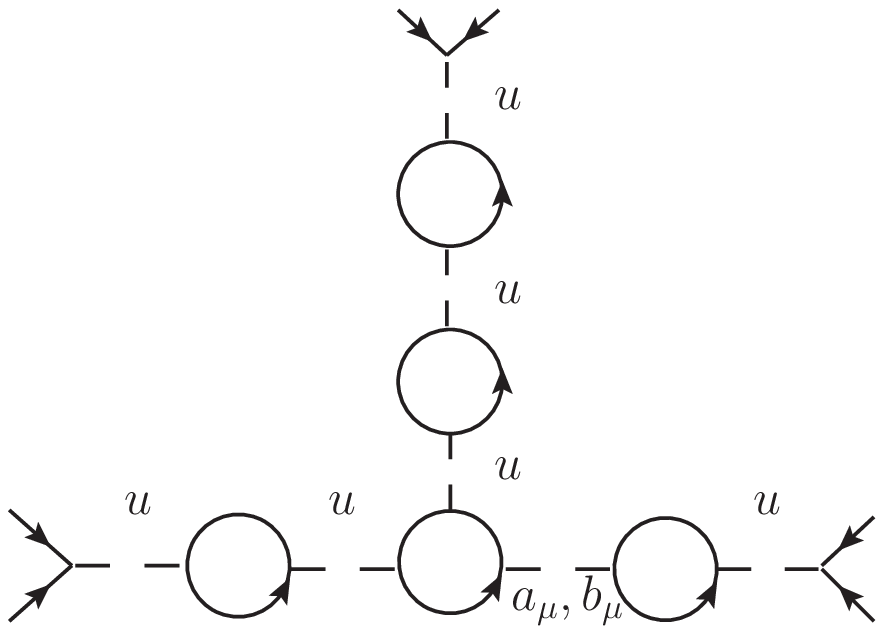}\hfill \includegraphics[width=.32\textwidth]{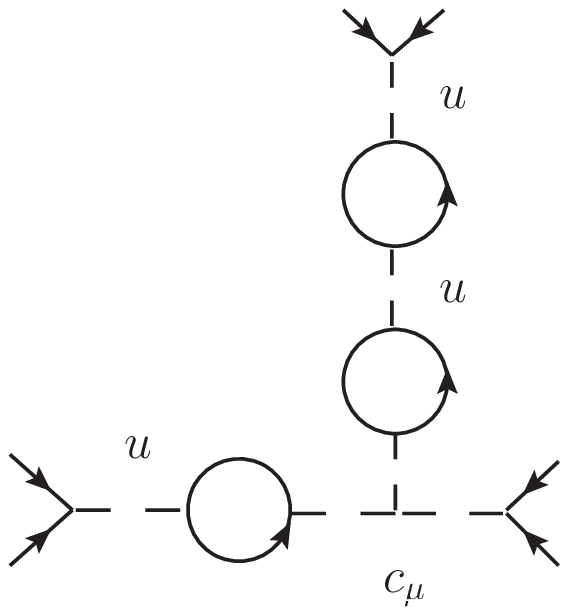}
			\end{center}
			\caption{Left: a diagram contributing to $\Gamma^{\left(6\right)}$, linear in $a_\mu$ or $b_\mu$. Right: a diagram contributing to $\Gamma^{\left(6\right)}$ proportional to $c_\mu$.}
			\label{gamma6a}
		\end{figure}

		The ${\Gamma}^{\left(6\right)}$ vertex has corrections linear in
		$c^\Lambda_{\mu}$ to leading order in the $\frac{1}{N}$ expansion (i.e. to order $\frac{1}{N^2}$), but it may also have contributions coming from the types of diagrams shown in Fig.~\ref{gamma6a} where $a_\mu$ or $b_\mu$ is inserted at the core (i.e. the inner loop with three propagator) or, as before, attached to an external leg. However, these corrections do not contribute to the scaling dimensions of the operators under study because, as explained above, $c_\mu$ does not contribute to the flows of $a_\mu$ and $b\mu$ at leading order in $1/N$. We therefore do not compute these corrections.
		
		The diagrams to be computed are exceptionally simple since they have a $c_\mu$ at the core with no loop and then just chain of bubbles with $u$ perturbations, these are schematically shown in Fig.~\ref{gamma6a}.

		The diagram (proportional to $\delta_{i_{1}i_{2}}\delta_{i_{3}i_{4}}\delta_{i_{5}i_{6}}$) with a chain with $n_1$ couplings $\hat{u}^\Lambda$ attached to the external momentums $p_1$ and $p_2$, a chain with $n_2$ couplings $\hat{u}^\Lambda$ attached to the external momentums $p_3$ and $p_4$, a chain with $n_3$ couplings $\hat{u}^\Lambda$ attached to the external momentums $p_5$ and $p_6$ and one $\hat{c}^\Lambda_{\mu}$ at the core is equal to:
		
		\begin{align}
		\nonumber\frac{-i\hat{c}^\Lambda_{\mu}}{2N^2}&\left(\frac{-\hat{u}^\Lambda}{4!}\right)^{n_1}\left(\frac{-\hat{u}^\Lambda}{4!}\right)^{n_2}\left(\frac{-\hat{u}^\Lambda}{4!}\right)^{n_3}\frac{4^{n_1+n_2+n_3}(n_1+n_2+n_3)!C^{n_1+n_2+n_3+1}_{1}}{\left(n_1+n_2+n_3+1\right)!}\\\nonumber
		\times&\left(\Pi\left(p_{1}+p_{2}\right)\right)^{n_1}\left(\Pi\left(p_{3}+p_{4}\right)\right)^{n_2}\left(\Pi\left(p_{5}+p_{6}\right)\right)^{n_3}\\
		\times&\Big[\left(p_{1}+p_{2}\right)^{\mu}(p_{1}+ p_{2})^2+\left(p_{3}+p_{3}\right)^{\mu}(p_{3}+ p_{4})^2+\left(p_{5}+p_{6}\right)^{\mu}(p_{5}+ p_{6})^2\Big] + \text{perms.}
		\end{align}
		
		\subsection{Running couplings}
		
		A convenient way to deduce the scaling dimensions of the operators coupled to
		$\hat{a}^\Lambda_{\mu}$ and $\hat{b}^\Lambda_{\mu}$, is to introduce an
		infrared regulator in propagators:
		\begin{equation}
		\frac{1}{q^2+r^\Lambda}\to \frac{1}{q^2+r^\Lambda+R_k(q)}
		\end{equation}
		and study the running of the various couplings when varying the
		regulator. We thus define the renormalized couplings as:
		\begin{align}
		\Gamma_{i_1i_2}^{(2)}(0)&=r_k \delta_{i_1i_2}\nonumber \\
		\Gamma_{i_1i_2i_3i_4}^{(4)}(0,0,0)&=\frac{\hat{u}_k}{3N}\Big(\delta_{i_1i_2}\delta_{i_3i_4}+\delta_{i_1i_3}\delta_{i_2i_4}+\delta_{i_1i_3}\delta_{i_2i_4}\Big)\nonumber \\
		\nonumber\Gamma^{\left(4\right), \mathcal O(p^3)}_{i_1i_2i_3i_4}(p_1,p_2,p_3)
		&=i\frac{\delta_{i_1i_2}\delta_{i_3i_4}}{2N}
		\bigg\lbrace (\hat{a}^k_{\mu}-\hat{b}^k_{\mu})\left(p_{1}+p_{2}\right)^{\mu}\left[p_{1} \cdot p_{2} - p_{3} \cdot p_{4}\right] 
		\\\nonumber&\qquad\qquad
		-\hat{b}^k_{\mu}\left(p_{1}^{\mu}p_{1}^2+p_{2}^{\mu}p_{2}^2+p_{3}^{\mu}p_{3}^2+p_{4}^{\mu}p_{4}^2\right)\bigg\rbrace
		+\text{2 perms}\\ \nonumber\Gamma^{\left(6\right), \mathcal O(p^3)}_{i_1i_2i_3i_4i_5i_6}(p_1,p_2,p_3,p_4,p_5)
		&=-i\frac{\delta_{i_1i_2}\delta_{i_3i_4}\delta_{i_5i_6}\hat{c}^k_{\mu}}{2N^2}
		\bigg\lbrace \left(p_{1}+p_{2}\right)^{\mu}(p_{1}+ p_{2})^2\\&+\left(p_{3}+p_{3}\right)^{\mu}(p_{3}+ p_{4})^2+\left(p_{5}+p_{6}\right)^{\mu}(p_{5}+ p_{6})^2\bigg\rbrace
		+\text{14 perms.}
		\end{align}
		One can then conclude that the running couplings are:
		\begin{align}
		\nonumber u_k=&\frac{u}{1+\frac{u_\Lambda}{6}\Pi_k\left(0\right)},\qquad a^\mu_k=\frac{a^\mu_\Lambda}{1+\frac{u_\Lambda}{6}\Pi_k\left(0\right)},
		\qquad b^\mu_k=\frac{b^\mu_\Lambda}{1+\frac{u_\Lambda}{6}\Pi_k\left(0\right)}\\& c^\mu_k=\frac{c^\mu_\Lambda}{\big(1+\frac{u_\Lambda}{6}\Pi_k\left(0\right)\big)^3}+Y_a a^\mu_\Lambda+Y_b b^\mu_\Lambda ,\qquad r_k=r^{\Lambda}+\overline{\Sigma}_{k}
		\end{align}
		where the functions $\overline{\Sigma}_{k}$ and $\Pi_{k}$ are calculated with the introduction of the infrared regulator:
		\begin{align}
		\overline{\Sigma}_{k}\left(r_\Lambda\right)&=\frac{u_\Lambda}{6}\int_q\frac{1}{q^2+r_k+R_{k}\left(q^2\right)}\\
		\Pi_{k}\left(0\right)&=\int_q\frac{1}{\left(q^2+r_k+R_{k}\left(q^2\right)\right)^{2}}.
		\end{align}
		Taking this into account, we obtain the flow of the running couplings:
		\begin{align}
		\partial_{t}r_k&=-\frac{\hat{u}_k}{6}\int_q \partial_{t}R_{k}(q) G_k^2(q)\nonumber\\
		\partial_{t}\hat{u}_k&=-\frac{\hat{u}_k^3}{18}\int_q  G_k^3(q)\int_{q'}\partial_{t}R_{k}(q') G_k^2(q') +\frac{\hat{u}_k^2}{3}\int_q \partial_{t}R_{k}(q) G_k^3(q)\nonumber\\
		\partial_{t}\hat{a}_k^{\mu}&=-\frac{\hat{a}_k^\mu \hat{u}_k^2}{18}\int_q G_k^3(q)\int_{q'}\partial_{t}R_{k}(q') G_k^2(q')
		+\frac{\hat{a}_k^{\mu}\hat{u}_k}{3}\int_q \partial_{t}R_{k}(q) G_k^3(q)\nonumber\\\nonumber
		\partial_{t}\hat{b}_k^{\mu}&=-\frac{\hat{b}_k^\mu \hat{u}_k^2}{18}\int_q G_k^3(q)\int_{q'}\partial_{t}R_{k}(q') G_k^2(q')
		+\frac{\hat{b}_k^{\mu}\hat{u}_k}{3}\int_q \partial_{t}R_{k}(q) G_k^3(q)\\\nonumber
		\partial_{t}\hat{c}_k^{\mu}&=-\frac{\hat{c}_k^\mu \hat{u}_k^2}{6}\int_q G_k^3(q)\int_{q'}\partial_{t}R_{k}(q') G_k^2(q')
		+\hat{c}_k^{\mu}\hat{u}_k\int_q \partial_{t}R_{k}(q) G_k^3(q)\\&+X_a \hat{a}_k^{\mu}+X_b \hat{b}_k^{\mu}
		\end{align}
		where, in the previous equations,
		$$ G_k(q)= \frac{1}{q^2+r_k+R_{k}\left(q^2\right)}.$$
		Introducing dimensionless and renormalized variables as explained in
		the main text, we retrieve the flow equations (\ref{eq:feN}).
		
		\section{Numerical method}\label{Ap:NumParam}

		We describe in this section the details of the numerical method used
		to detemine the scaling dimensions in the $\mathcal O(\partial^3)$
		approximation of NPRG.
		
		We first determine the fixed point of the $O(N)$ model at
		$\mathcal{O}(\partial^2)$ in a $\tilde\rho$ grid with $N_\rho+1$ sites in a
		box $\tilde\rho \in [0,L_\rho]$ (this corresponds to a step in the
		$\tilde\rho-$lattice $\Delta\rho=L_\rho/N_\rho$). The derivatives are
		approximated by centered finite differences with five points (with the
		exception of the edges, i.e. the first two and last two sites, where
		we use lateral finite differences).
		
		The internal momentum integrals that appear are one dimensional (the
		angular part is just a constant) and are calculated by a
		Legendre-Gauss quadrature with $N_q$ points in a box of size
		$|L_q|\equiv \frac{q_{max}}{k}$.
		
		The normalization condition is fixed as
		$\tilde{Z}(\tilde{\rho}_i)|_{i=N_\rho/3}=1$, where
		$\tilde{Z}(\tilde{\rho})$ is the dimensionless version of $Z_k(\rho)$
		and $\tilde{\rho}_i$ is the value of $\tilde{\rho}$ at site $i$. On
		top of this, the size of the box $L_\rho$ is adjusted every time a new
		set of parameter is considered so that the minimum of the potential at
		the fixed point falls in the site $i=N_\rho/3$.
		
		The parameters were chosen so as that the numerical error in the leading exponents ($\eta$ and $\nu$) are
		is below one per mille in $d=3$. Then we varied the values of $d$ for each $N$. In particular we used
		the exponential regulator $R_{k}(q^2)=\mathcal{Z}_{k}q^2 r(q^2/k^2)$ with:
		\begin{equation}
		\label{expreg}
		r(y)=\frac{\alpha}{e^{y}-1}
		\end{equation}
		where $\mathcal{Z}_k$ is the field renormalization which is related to
		the running anomalous dimension by
		$\partial_t\mathcal{Z}_k=-\eta_k\mathcal{Z}_k$ (when approaching the
		fixed point $\eta_k$ approaches the field anomalous dimension). The
		parameter $\alpha$ was fixed in order to estimate the dependence of
		the results on the arbitrary regulating function $R_k(q)$. In order to
		do so, for each $N$ we employ the criterium of minimal sensitivity
		(PMS) \cite{Stevenson:1981vj}. That is, we choose the value of
		$\alpha$ which minimizes the dependence on the regulator of some
		observable by varying the regulator in the family \eqref{expreg}.
		This procedure has been tested before and shown to be very effective
		and predictive \cite{Canet:2002gs,Balog:2019rrg}. In the present implementation we
		choose $\alpha$ to be an extremum of the critical exponent $\eta$ for
		$d=3$ (no significative difference was observed when doing PMS on
		another quantity such as critical exponent $\nu$).
		
		After obtaining the fixed point, we computed the eigenvectors of the
		$3N_\rho$ linear system ($N_\rho$ variables for each function
		$\tilde{a}_{\mu}$, $\tilde{b}_{\mu}$ and $\tilde{c}_{\mu}$) at the
		fixed points determined previously. To solve numerically the
		$\mathcal{O}(\partial^3)$ of the DE for the $O(N)$ model the following
		set of parameters was used:
		\begin{equation}
		N_\rho=60,\qquad L_\rho=3\rho_0,\qquad N_q=15,\qquad L_q=4.2.
		\end{equation}

		\section{Inequalities}\label{inequalities}
		
		In the present Appendix, we prove by induction the inequality \eqref{eq:sufCond}
		for $T\ge T_c$.  We use generalizations of Griffiths and Lebowitz
		inequalities which were proven for $N=2,3,$ and $4$
		\cite{bricmont1977gaussian,dunlop1979zeros,dunlop1976correlation,monroe1979correlation,kunz1975correlation,sokal1982mean}. The
		inequalities obtained in those references are the following:
		\begin{align}
		&\langle \phi_1(x^{(1)}_1)\dots \phi_1(x^{(1)}_{n_1})\phi_2(x^{(2)}_1)\dots \phi_2(x^{(2)}_{n_2})\dots \phi_N(x^{(N)}_1)\dots \phi_N(x^{(N)}_{n_N})\rangle\ge 0, \label{G1}\\
		&\langle \phi_\alpha(x_1)\dots \phi_\alpha(x_n)\phi_\alpha(y_1)\dots \phi_\alpha(y_m)\rangle\ge 
		\langle \phi_\alpha(x_1)\dots \phi_\alpha(x_n)\rangle\langle \phi_\alpha(y_1)\dots \phi_\alpha(y_m)\rangle,  \label{G2}\\
		&\langle \phi_\alpha(x_1)\dots \phi_\alpha(x_n)\phi_\beta(y_1)\dots \phi_\beta(y_m)\rangle\le 
		\langle \phi_\alpha(x_1)\dots \phi_\alpha(x_n)\rangle\langle \phi_\beta(y_1)\dots \phi_\beta(y_m)\rangle \,\mathrm{with}\,\alpha\neq\beta.  \label{Lebowitz}
		\end{align}
		Inequality \eqref{G1} and \eqref{G2}
		are very similar to the Griffiths  inequalities I and II for a scalar field and \eqref{Lebowitz} is
		very similar to the Lebowitz inequality.
		
		First of all, it is clear that for $T\ge T_c$ we need only to consider
		correlations with $n$ and $m$ with the same parity.
		In order to begin the induction, we observe that the inequality is trivially true for $n+m=2$. We assume now the validity of \eqref{eq:sufCond} for all
		values of $m$ and $n$ such that $m+n<N$. Under this hypothesis, we then prove its validity for all $\{m,n\}$ such that $m+n=N$.
		
		We first consider the case where $n$ and $m$ are odd.
		This case is simpler because the second term of the l.h.s of \eqref{eq:sufCond} is zero because we are considering
		$T\ge T_{c}$ and \eqref{G1} readily shows that $\nonumber G_{m,n}^{odd}\geq 0$.
		By symmetry, the structure of the two point function must take the form:
		\begin{align}
		\nonumber G_{m,n}^{odd}\left(x,y\right)\equiv\left\langle \varphi_{i_1}\left(x\right) \cdots \varphi_{i_m} \left(x\right)\right.&\left.\varphi_{j_1}\left(y\right)\cdots  \varphi_{j_n}\left(y\right)\right\rangle =   \\
		\nonumber \sum_{l=0}^{\frac{n-1}{2}}f_{l}\left(x,y\right)\Big[\delta_{i_{1}j_{1}}\cdots\delta_{i_{2l+1}j_{2l+1}} & \delta_{i_{2l+2}i_{2l+3}}\delta_{j_{2l+2}j_{2l+3}}\cdots\delta_{i_{m-1}i_{m}}\delta_{j_{n-1}j_{n}}+\\
		&\left(\frac{m!n!}{\left(2l+1\right)!\left(m-2l-1\right)!!\left(n-2l-1\right)!!}-1\right)\text{perms.}\Big]\label{eq_decomp}
		\end{align}
		where, without loss of generality we focused on the case $m\geq n$. In the previous sum, $2l+1$ corresponds to the number of Kronecker delta which connect $i$'s with $j$'s.
		In the first step of the proof, we consider the following configuration of indices:
		\begin{align}
		\nonumber & i_{k}= \begin{cases}
		1 & \text{for\quad} k=1 \\ 2 &  \text{for\quad }   k=2,\cdots,m
		\end{cases} \\
		\nonumber & j_{k}= \begin{cases}
		2 & \text{for\quad} k=1,\cdots,2s  \\ 1 & \text{for\quad} k=2s+1,\cdots,n
		\end{cases}
		\end{align}
		with $s$ ranging from $0$ to $\frac{n-1}{2}$. (For $s=0$, all the $j_k$ are $1$.)
		We can now apply inequality \eqref{Lebowitz} for these index configurations to get:
		\begin{align}
		\nonumber \tilde G_{m,n}^{odd}\left(x,y\right)&\leq  \left\langle \varphi_{1}\left(x\right)\left(\varphi_{1}\left(y\right)\right)^{n-2s}\right\rangle  \left\langle \left(\varphi_{2}\left(x\right)\right)^{m-1}\left(\varphi_{2}\left(y\right)\right)^{2s}\right\rangle  \\
		& \leq CG\left(x-y\right)
		\end{align}
		where the tilde on $G$ indicates that it is taken in a particular configuration of indices and where we have used the recursion hypothesis \eqref{eq:sufCond} for $m'+n'<m+n$ and the fact that  $\left\langle \left(\varphi_{2}\left(x\right)\right)^{m-1}\left(\varphi_{2}\left(y\right)\right)^{2s}\right\rangle$ is bounded by a constant to obtain the last line.
		Using, on the other hand the decomposition \eqref{eq_decomp}, we get:
		\begin{align}
		\sum_{t=0}^{s} \frac{\left(2s\right)!\left(n-2s\right)!!\left(m-1\right)!}{\left(2t\right)!\left(2s-2t\right)!!\left(m-1-2t\right)!!}f_{t}\left(x,y\right)  \leq CG\left(x-y\right)
		\end{align}
		By considering the different possible values of $s$, we easily show that the $f_t$ are bounded by $G(x-y)$ (up to a multiplicative constant) which ensure that the inequality is valid for $G_{m,n}^{odd}$.

		The even case is a bit different since the second term of the l.h.s. of \eqref{eq:sufCond} is nonzero. We again choose $m\geq n$ and look at the structure of the two point function:
		\begin{align}
		\nonumber G_{m,n}^{even}\left(x,y\right)\equiv\left\langle \varphi_{i_1}\left(x\right) \cdots \varphi_{i_m} \left(x\right)\right.&\left.\varphi_{j_1}\left(y\right)\cdots  \varphi_{j_n}\left(y\right)\right\rangle = \\ \nonumber \sum_{l=0}^{\frac{n}{2}}g_{l}\left(x,y\right)\left(\delta_{i_{1}j_{1}}\cdots\delta_{i_{2l}j_{2l}} \right. & \left. \delta_{i_{2l+1}i_{2l+2}}\delta_{j_{2l+1}j_{2l+2}}\cdots\delta_{i_{m-1}i_{m}}\delta_{j_{n-1}j_{n}}\right. +\\
		& \left. \left(\frac{m!n!}{\left(2l\right)!\left(m-2l\right)!!\left(n-2l\right)!!}-1\right)\text{perms.}\right)\label{eq_g}
		\end{align}
		Inequality \eqref{G1} readily imposes that $G_{m,n}^{even}\left(x,y\right)\geq 0$. We now proceed in the same way as for the odd case. We take a configuration of indices:
		\begin{align}
		\nonumber & i_{k}= \begin{cases}
		1 & \text{for\quad} k=1 \\ 2 & \text{for\quad} k=2,\cdots,m \end{cases} \\
		\nonumber & j_{k}= \begin{cases}
		2 &\text{for\quad}  k=1,\cdots,2s-1  \\ 1 &\text{for\quad}  k=2s,\cdots,n
		\end{cases}
		\end{align}
		with s ranging from $1$ to $\frac{n}{2}$. These configurations in combination with the inequality \eqref{Lebowitz} impose an upper bound on a strict conical combinations of the $g_{i}$ functions \textit{not} involving the $g_{0}$ function:
		\begin{align}
		\nonumber  \tilde G_{m,n}^{even}\left(x,y\right)&=\sum_{t=1}^{t=s} \frac{\left(2s-1\right)!\left(n-2s+1\right)!!\left(m-1\right)!}{\left(2t-1\right)!\left(2s-2t\right)!!\left(m-2t\right)!!}g_{t}\left(x,y\right) \\
		\nonumber&\leq  \left\langle \varphi_{1}\left(x\right)\left(\varphi_{1}\left(y\right)\right)^{n-2s+1}\right\rangle  \left\langle \left(\varphi_{2}\left(x\right)\right)^{m-1}\left(\varphi_{2}\left(y\right)\right)^{2s-1}\right\rangle  \\
		& \leq CG^{2}\left(x-y\right)
		\end{align}
		where we have made use of the validity of recursion hypothesis \eqref{eq:sufCond} for all $m'+n'<m+n$ and the tilde is again used to recall that a particular configuration of indices was used. We, again, find a lower and an upper bound on a strict conical combination of the functions $g_{s}$, with $s=1,\cdots,\frac{n}{2}$. This implies that the absolute value of each of these $g_{s}$ functions (with $s\neq 0$) is bounded by a constant times $G^{2}\left(x-y\right)$ (by arguments identical to the odd case).
		To complete the argument we need to study the function $g_{0}$. This one is clearly particular because it corresponds to the case where no $i$ or $j$ are connected through a Kronecker delta (see Eq.~\eqref{eq_decomp}). To do this we consider the even simpler configuration where all indices $i=1$ and all $j=2$, this yields for the two-point function:
		\begin{equation}
		0 \leq G_{m,n}^{even}\left(x,y\right)=g_{0}\left(x,y\right)\left(m-1\right)!!\left(n-1\right)!!\leq \left\langle\varphi^{m}_{1}\left(0\right) \right\rangle \left\langle\varphi^{n}_{1}\left(0\right) \right\rangle
		\end{equation}
		from which immediately follows that:
		\begin{equation}
		\label{eq:EvLowBo}
		g_{0}\left(x,y\right)-\frac{\left\langle\varphi^{m}_{1}\left(0\right) \right\rangle \left\langle\varphi^{n}_{1}\left(0\right) \right\rangle}{\left(m-1\right)!!\left( n-1\right)!!} \leq 0
		\end{equation}
		
		Let's consider now all indices $i,j=1$ to obtain a lower bound on a strict conical combination of \textit{all} the $g$ functions by making use of the inequality \eqref{G2}:
		\begin{equation}
		\sum_{l=0}^{\frac{n}{2}}\frac{m!n!}{\left(2l\right)!\left(m-2l\right)!!\left(n-2l\right)!!}g_{l}\left(x,y\right)\geq \left\langle\varphi^{m}_{1}\left(0\right) \right\rangle \left\langle\varphi^{n}_{1}\left(0\right) \right\rangle
		\end{equation}
		and combining this with \eqref{eq:EvLowBo} we obtain a lower bound:
		
		\begin{equation}
		\begin{split}
		g_{0}\left(x,y\right)-\frac{\left\langle\varphi^{m}_{1}\left(0\right) \right\rangle \left\langle\varphi^{n}_{1}\left(0\right)
			\right\rangle}{\left(m-1\right)!!\left( n-1\right)!!} & \geq -\frac{1}{\left(m-1\right)!!\left( n-1\right)!!}\sum_{l=1}^{\frac{n}{2}}\frac{m!n!}{\left(2l\right)!\left(m-2l\right)!!\left(n-2l\right)!!}g_{l}\left(x,y\right)\\&\geq C G^{2}\left(x-y\right)
		\end{split}    
		\end{equation}
		with a constant $C<0$.
		
		So, we have bounded all the $g$ functions with the exception of $g_{0}$ for which the bound involves $g_{0}$ minus a constant. It turns out that this constant has a simple interpretation: when inserted in Eq.~\eqref{eq_g}, the multiplicative constants simplify and we are just left with $\left\langle\varphi^{m}_{1}\left(0\right) \right\rangle \left\langle\varphi^{n}_{1}\left(0\right) \right\rangle$, which exactly compensates the diconnected part appearing in~\eqref{eq:sufCond}. This concludes the proof of~\eqref{eq:sufCond}.

	\end{appendix}
	
	\bibliographystyle{spphys}
	\bibliography{confON_v3}
	
\end{document}